\documentclass[12pt]{article}

\usepackage{times}
\usepackage{changepage} 
\usepackage{hyperref}
\usepackage[T1]{fontenc} 
\usepackage{breakcites} 
\usepackage{url} 

\newcommand{\itemb}{\begin{itemize}}
\newcommand{\iteme}{\end{itemize}}

\newcommand{\enumb}{\begin{enumerate}}
\newcommand{\enume}{\end{enumerate}}

\newcommand{\smallest}[1]{\vspace{6pt}\noindent {\bf  {#1}.}}
 
\newcommand{\smallestq}[1]{\vspace{6pt}\noindent {\bf #1?}}

\newcommand{\com}[1]{}
\usepackage{xcolor}

\newcommand{\crh}{corticotropin-releasing hormone}

\newcommand{\acth}{adrenocorticotropic hormone}

\newcommand{\testos}{testosterone}
\newcommand{\Testos}{Testosterone}
\newcommand{\dht}{dihydrotestosterone}

\newcommand{\dhea}{dehydroepiandrosterone}

\newcommand{\oxt}{oxytocin}

\newcommand{\asd}{{{autism spectrum disorder}}}
\newcommand{\Asd}{{{Autism spectrum disorder}}}
\newcommand{\pcos}{polycystic ovary syndrome}

\newcommand{\differenti}{differentiation}

\newcommand{\xtblty}{excitability}

\newcommand{\epig}{epigenetic}

\newcommand{\hypog}{hypoglycemia}

\newcommand{\hyperg}{hyperglycemia}

\newcommand{\desensit}{desensitization}

\newcommand{\resist}{resistance}

\newcommand{\metab}{metabolism}

\newcommand{\hthal}{hypothalamus}

\newcommand{\amg}{amygdala}

\newcommand{\hippo}{hippocampus}

\newcommand{\bnst}{bed nucleus of the stria terminalis}

\newcommand{\icell}{{{intracellular}}}

\newcommand{\sympa}{{{sympathetic}}}

\newcommand{\infl}{{{inflammation}}}

\newcommand{\plast}{{{plasticity}}}

\newcommand{\catwo}{{{Ca}\textsuperscript{{\raisebox{-1pt}{$\scriptstyle{2+}$}}}}}

\newcommand{\thn}{{T*ASD}} 

\usepackage{fancyhdr}
\title{A CRH Theory of Autism Spectrum Disorder}
\author{
{Ari Rappoport}\\ 
The Hebrew University of Jerusalem, Israel\\
ari.rappoport@mail.huji.ac.il
}
\date{June 2024}
\begin{document}
\maketitle
\begin{adjustwidth}{30pt}{30pt}
{\bf Abstract.} 
This paper presents a complete theory of \asd\ (ASD), explaining its etiology, symptoms, and pathology. The core cause of ASD is excessive stress-induced postnatal release of \crh\ (CRH). CRH competes with urocortins for binding to the CRH2 receptor, impairing their essential function in the utilization of glucose for growth. This results in impaired development of all brain areas depending on CRH2, including areas that are central in social development and eye gaze learning, and low-level sensory areas. Excessive CRH also induces excessive release of adrenal androgens (mainly DHEA), which impairs the long-term \plast\ function of gonadal steroids. I show that these two effects can explain all of the known symptoms and properties of ASD. The theory is supported by strong diverse evidence, and points to very early detection biomarkers and preventive pharmaceutical treatments, one of which seems to be very promising. 

\end{adjustwidth}
\section{Introduction}
\Asd\ (ASD) is a developmental disorder defined by impaired social communication and restricted, repetitive behaviors \cite{lord2020aut, hirota2023aut}. 
The average global prevalence of ASD is around 1\%, higher in high-income countries \cite{lord2020aut}. 
Prevalence in 8 year olds in the US in 2020 was reported to be much higher, approximately 4\% of boys and 1\% of girls \cite{maenner2023pre}. 

Despite the high public profile of ASD, the extreme burden it puts on patients and caregivers, and extensive research efforts (over 77K/2.2M papers in pubmed/Google Scholar), there is currently no theory of ASD, no early biomarkers, and no preventive treatment. Post-diagnosis treatment is only behavioral (no pharmaceutical treatment), and its efficacy is limited. 


Here I present a {\bf CRH theory of ASD (\thn)}. \thn\ is the first complete theory of ASD, explaining its etiology, symptoms, pathophysiology, and treatment. The theory is supported by substantial evidence, and points to novel very early screening and treatment procedures. 

\smallest{Theory overview}
According to \thn, the core cause of ASD is pre-, peri- or postnatal stress that induces excessive postnatal corticotropin-releasing hormone (CRH) release. The main trigger is a stress-inducing state, but the tendency for excessive stress-induced CRH can be heritable. Excessive release can start prenatally, but the ASD symptoms are mainly due to its postnatal effects. The damage is done during the first year of life, mainly at 3-9 months. 

Excessive postnatal CRH impairs normal brain development. This happens via two arms, urocortins and steroids (STRDs). The common aspect of the two arms is prolonged stress-associated signaling at the expense of long-term \plast\ signaling. 

Urocortins (UCNs) are members of the CRH peptide family, which bind CRH receptor 2 (CRH2). UCNs are released in high glucose states, and are crucial for the utilization of glucose for growth (protein synthesis). In ASD, the excessive CRH yields prolonged binding of CRH2 by CRH instead of UCNs, which impairs the long-term growth effects of CRH2 signaling. 

In the second damage arm, excessive CRH induces excessive ACTH, which (in addition to inducing cortisol) is the main inducer of adrenal \dhea\ (DHEA), a precursor of \testos\ and estrogen. STRDs act via nuclear receptors to support long-term \plast\ (\differenti, protein synthesis). However, they also support the initial stages of \plast\ via membrane receptors. These two paths are in mutual opposition. DHEA supports rapid signaling, and its excess in ASD impairs the long-term \plast\ aspects of gonadal STRDs. 

The UCN and STRD arms can explain all of the major ASD symptoms and properties. Impaired social interaction, increased sensory sensitivity (including aversion to touch), impaired eye gaze, and early motor impairment are a consequence of the high CRH2/UCN expression in social, low-level sensory, eye control, and motor brain areas. 
Impaired social interaction and touch aversion are also explained by the suppression of oxytocin (OXT), an agent essential for social development and pleasant touch, by \testos\ and CRH. 
Repetitive stereotypies and self-injurious behavior such as scratching are directly stimulated by CRH and ACTH.

Limited interests and aversion to novelty are present in ASD because novelty induces CRH release, which people with ASD seek to minimize. These behaviors are a form of self-treatment. 

Males are more vulnerable to ASD for at least three partially overlapping reasons. First, estrogen directly supports glucose metabolism and protein synthesis. Second, males need postnatal STRD-mediated \plast\ more than females do, as attested by higher release during minipuberty. Third, estrogen stimulates the release of OXT, while \testos\ opposes it. 

Evidence supporting \thn\ includes direct evidence for increased CRH and ACTH and their effects (e.g., on the \sympa\ nervous system and mast cells), and data related to anatomical growth, growth pathways, STRDs, OXT, food intake patterns, comorbidities, and sexual characteristics. 

Blood and urine tests for the agents involved in the impaired paths (CRH, ACTH, DHEA etc) during the first postnatal months, and adrenal imaging, can serve for very early identification. There are various possible preventive treatments, probably the most promising being transdermal DHT patches, which reduce CRH and ACTH release. 

The paper details the theory, its supporting evidence, and possible treatments. 

\section{Theory}

The core cause of ASD is pre-, peri- or postnatal stress that induces excessive postnatal CRH release. The stress event itself can be minor and/or chronic, with the response enhanced by a tendency for increased release. A large part of the genome supports stress sensing and responses, so such a tendency can involve a large number of small genetic and epigenetic changes. The excessive response can start prenatally, but the ASD symptoms are mainly due to damage done during the first postnatal year (say, 3 to 9 months). The excessive stress responses usually subside after the first year to normal or only slightly increased levels, but their effects persist. Thus, fundamentally, ASD is an `injury' type disorder. 

Excessive postnatal CRH impairs normal brain development via two arms, UCNs and STRDs. The two arms share a common aspect, prolonged stress-associated signaling of an agent at the expense of its long-term \plast\ signaling. 

\subsection{Urocortins (UCNs)}

 The first damage arm involves {\bf urocortins (UCNs)}, members of the CRH peptide family
 \cite{bale2004crf, deussing2018cor}.  
CRH acts on two receptors, CRH1 and CRH2, with higher affinity for CRH1. UCN1 binds these two receptors as well, while UCN2 and UCN3 bind only CRH2. The two receptors are G protein-coupled receptors that mainly bind Gs to induce cAMP, but also bind Gi/o and Gq.

{\bf The role of CRH1 is to trigger responses to metabolic deficiency (\bf hypo) stress states}, including \hypog\ (hypog), cold, and hypoxia. It triggers the hypothalamus (hthal)-pituitary-adrenal (HPA) stress axis to yield ACTH and cortisol release, stimulates the counter-regulatory responses (CRRs) to hypog via the \sympa\ nervous system (SNS), acts in the hthal to increase food intake, and promotes general neural excitability. 

 Regarding CRH2, at present there is no coherent view of its action, besides a general opposition to some of the effects of CRH1  \cite{bale2004crf, deussing2018cor}.   \thn\ provides a novel coherent account, distinguishing between CRH2 binding by UCNs ({\bf U-CRH2}) and by CRH ({\bf C-CRH2}). 

\thn\ presents a new theory of U-CRH2, whereby its role is to  {\bf trigger responses to metabolic excess (hyper) states}, mainly \hyperg\ ({\bf hyperg}), and also heat. In support of this theory, we note that UCNs are released in unstressful situations in response to high meal-induced glucose \cite{li2003uro, li2007uro, simpson2020ucn}, 
and that U-CRH2 supports glucose uptake by tissues, insulin release, and glucose utilization for growth (protein synthesis, differentiation)
\cite{chanalaris2005hyp, markovic2011mol, huang2012his, roustit2014uro, walther2014uro, gao2016one, kavalakatt2021uro, meister2022viv, lautherbach2022uro, flaherty2023chr}.
 U-CRH2 opposes CRRs \cite{mccrimmon2006cor}, 
and suppresses appetite and food intake \cite{bale2004crf}.

 Due to the low CRH affinity to CRH2, {\bf C-CRH2 is activated in prolonged and/or high stress situations}. 
While U-CRH2 induces quick CRH2 endocytosis and recycling back to the plasma membrane \cite{hauger2013des}, 
 C-CRH2 delays endocytosis \cite{hauger2013des}, opposes glucose uptake and insulin signaling \cite{hauger2013des}, 
 and promotes prolonged cAMP production \cite{markovic2011mol}. 
As a result, {\bf C-CRH2 opposes the growth effects of UCN signaling}.

 {\bf In ASD, chronic CRH release yields chronic postnatal C-CRH2.} This reduces both U-CRH2 binding (because CRH competes with UCNs for CRH2 binding) and U-CRH2 signaling, impairing the learning of all that should be learned during these crucial months. Chronic CRH may also involve chronic activation and \desensit\ of CRH1. 

 UCNs are crucial for the development of areas showing CRH2 expression and UCN release. These include the \amg, BNST, lateral septum, visual cortex, the Edinger-Westphal (EW) nucleus, \hthal, auditory cortex, olfactory and touch circuits, motor circuits, and prefrontal cortex (PFC). These areas are those that support low-level sensory inputs (olfaction, vision, audition, touch), sexual and social development, executive function, and regulation of metabolism. 

 UCNs are also expressed in non-brain tissue, and there are indeed widespread general health problems in ASD \cite{croen2015hea}. 
 However, the brain is more sensitive to UCN impairment, because in non-brain tissue glucose-induced growth is strongly supported by insulin. In the brain, insulin mainly acts on motor pathways. 

\subsection{Steroids (STRDs)}

The second damage arm in ASD involves {\bf sex steroids (STRDs)}. Excessive CRH induces excessive ACTH, which is known as the inducer of adrenal glucocorticoid (GC, cortisol in humans) release. However, {\bf ACTH is also the main inducer of adrenal \dhea\ (DHEA)} \cite{turcu2014adr},
 the precursor of \testos\ and estrogen. 
At the same time, high CRH reduces GSTRDs via CRH2 \cite{kageyama2013reg}. 



 All steroids, including gonadal STRDs and GC, are known to act via nuclear receptors to support long-term \plast. However, they also support the initial stages of \plast\ via membrane receptors that induce rapid \icell\ signaling \cite{norman2004ste}. 
 These two paths oppose each other,  because the \catwo, cAMP and MAPK signaling activated during rapid initial responses interfere with the safe DNA access needed for long-term \plast. DHEA generally supports rapid signaling, as shown by its direct effects on \catwo\ \cite{clark2018mec} and by its mutual opposition with GC \cite{saponaro2007lon, mcnelis2013deh, clark2018mec}. 
 Thus, DHEA and adrenal steroids (ASTRDs) generally oppose the long-term \plast\ aspects of gonadal steroids (GSTRDs). 

During the first 1-6 months after birth, human babies undergo {\bf minipuberty}, very high GSTRD release, indicating that GSTRD-mediated \plast\ is crucial for early postnatal development. In parallel, the adrenal gland undergoes involution and loses about half of its volume \cite{ben2007par, dhayat2017and}. 
In ASD, these processes are impaired, and {\bf the excessive postnatal ASTRDs impair GSTRD-mediated development}. 


%

\subsection{Explaining ASD symptoms and properties}

 The UCN and STRD arms of ASD can explain all of the major ASD symptoms and properties, as follows.

 {\bf Impaired social interaction} is a defining characteristic of ASD, explained in two ways. First, there is high CRH2 expression in the areas that support social development, including the BNST, \amg, lateral septum, and \hthal. This indicates that these areas need U-CRH2 signaling, which is impaired. 
 
 The second account is related to {\bf oxytocin (OXT)}. 
The OXT system is one of the most important systems in early postnatal development, when human babies learn that the company of other humans is a positive thing via its association with food and pleasant touch. The role of OXT as an agent essential for the development of sociality is well accepted \cite{carter2014oxy}. 
UCN3 greatly enhances OXT-induced accessory olfactory bulb \plast\ \cite{frankiensztajn2018com}. 
During the first few years after birth, three is an almost complete overlap between UCN and OXT expression in the \hthal\ PVH and supraoptic nuclei \cite{ohno2018inc}. 
OXT is directly inhibited by both \testos\ \cite{dai2017dir} and C-CRH2 \cite{chu2013eff, bosch2016oxy, martinon2018cor}. 
These OXT-UCN data further explain the social impairment in ASD. 

 {\bf Repetitive stereotypies} (arm flapping, hand twirling, clapping, rocking, head banging) are a defining characteristic of ASD. CRH \cite{matsuzaki1989eff, strome2002int, bakshi2007sti, carpenter2007cor, ogino2014com} and ACTH \cite{vanErp1993dif, wikberg2000new} directly induce these types of movements in animal models, including non-human primates, 
probably  because they dramatically increase neural \xtblty. (See also itch below.)

 {\bf Limited interests, extreme aversion to novelty, and a desire for order} are defining characteristics of ASD. CRH release, which is already basally excessive in ASD, is acutely stimulated by unpredicted inputs \cite{sherman1986icv, gesing2001psy}, and makes male goats show more distress following novelty \cite{roussel2005gen}. 
 Conversely, medial \amg\ UCN3 increases preference for novel conspecifics \cite{shemesh2016ucn}. 
 People with ASD reduce the unpleasant feeling associated with additional CRH release by restricting novel inputs. In addition, brain learning is impaired in ASD, so people with ASD naturally tend to engage in actions and thoughts that they have already learned.

 {\bf Increased sensory sensitivity} is a known property of people with ASD. This happens both because CRH1 promotes sensory hyper\xtblty, and because CRH2 is highly expressed in low-level sensory paths. Since early postnatal development mainly involves pruning of excitatory synapses and maturation of inhibition \cite{takesian2013bal, quast2017dev}, its impairment leads to sustained sensory overload. 

 {\bf Self-injurious behavior} is very common in ASD \cite{steenfeldt2020pre} (some stereotypies involve self-injury as well). 
 ACTH, aMSH, and beta-endorphin (other products of the ACTH gene POMC) induce itch and scratching \cite{vanErp1993dif, yamamoto2010inv}, 
 and both ACTH \cite{fetissov2006agg, veenema2007low} and DHEA stimulate aggression \cite{pajer2006adr, soma2015dhe}. 
%

 {\bf Aversion to touch} is an early sign of ASD. This can be explained by sensory hypersensitivity, ACTH-mediated itch, and reduced OXT, which mediates pleasant touch \cite{mcglone2014dis}. 
The excessive stimulation of the itch path induces an allodynia-like state in which neutral inputs evoke aversive itch signaling. 

 {\bf Impaired eye gaze} is a known marker of ASD. The EW nucleus controls learned eye gaze \cite{kozicz2011edi} and contains a population of centrally-projecting neurons with very high UCN1 expression \cite{kozicz2011edi}. 
EW UCN1 is not expressed at birth and gradually increases, showing the importance of UCN signaling to EW development \cite{cservenka2010pos}. 

 {\bf Early motor impairment} is clearly present in ASD \cite{lim2021ear}. 
CRH2 is abundantly expressed in skeletal muscle neurons \cite{samuelsson2004cor}, 
and UCNs act via CRH2 to improve muscle metabolism and growth \cite{hinkle2003act, jamieson2011uro, reutenauer2012uro}. 
STRDs are known to be important for motor development. 

 {\bf Males are more vulnerable} to ASD for at least three reasons. First, estrogen, which is higher in females, directly supports brain glucose metabolism and protein synthesis \cite{azcoitia2019mol}.  Thus, the relative female advantage in estrogen covers up for some of the damage done in ASD. 
Second, minipuberty occurs in both sexes but is much stronger in males \cite{kuiri2014act}. This shows that males need postnatal GSTRD-mediated \plast\ more than females do. 
Relatedly, EW UCN1 expression is 10-1.6 higher in males than in females \cite{derks2010sex}, making male eye gaze learning more sensitive to excessive CRH. 
Third, estrogen stimulates the release of the social agent OXT \cite{sharma2012erb, hiroi2013and}, while \testos\ opposes it \cite{dai2017dir}. 

The {\bf spectrum nature} of ASD stems from the wide expression pattern of CRH, UCNs, CRH receptors, and STRD receptors, and the fact that the core cause is stress, which can be highly variable, spatially and temporally. Hence, there is large variability in the ways in which the damage can be manifested. 

 
 It is common for babies with ASD to show {\bf excessive passivity} during their first year of life, followed by {\bf hyperactivity} during their second year (and throughout life) \cite{zwaigenbaum2005beh}. 
 This datum supports high CRH during the 1st year, which activates CRH2 to suppress hunger, cortical norepinephrine, and the SNS. CRH release gradually diminishes, until at some point it can chronically activate CRH1, promoting hypermovement, hunger, food intake, gut issues, etc.
 Indeed, early life stress has been shown to induce permanent CRH hyper\xtblty\ \cite{hu2020ear}. 

Although the public image of {\bf the autistic savant} is clearly unfounded, a small number of papers reported that many people with ASD excel in specific limited areas (e.g., mathematics, visual perception) \cite{howlin2009sav, meilleur2015pre}. 
In addition, ASD risk is higher with parents with higher IQ and technical capabilities \cite{gardner2020ass}. 
These data can be explained by a larger number of synapses (especially excitatory ones) due to decreased pruning during postnatal development, and by chronic CRH stimulation of the release of norepinephrine, which promotes memory \cite{roozendaal2011mem}. 
Conversely, {\bf intellectual impairment} is highly comorbid with ASD, and is explained by reduced brain (specifically PFC) development. 

Cortisol exerts negative feedback on CRH and ACTH secretion. Since STRDs (mainly ASTRDs) oppose GC \cite{seale2004gon}, 
\thn\ predicts that any situation in which both CRH and ASTRDs (and GSTRDs) are high during development should increase ASD risk. People are especially vulnerable in periods with normally high GSTRDs, because they need their long-term \plast\ effects in these periods. This is indeed the case (see PCOS and stress during pregnancy below).

\section{Evidence}
\thn\ is supported by extensive evidence. The main lines are related to stress, CRH, ACTH, growth (anatomy and pathways), STRDs, OXT, comorbidities, and food intake. 

\subsection {Stress, CRH, ACTH}
There is strong evidence for increased CRH and ACTH, relating to the prenatal, perinatal, and post-damage periods. 

\smallest{Perinatal and prenatal stress} 
Perinatal stress, which involves high CRH release, increases ASD risk \cite{gardener2011per}. This includes complications during delivery \cite{gardener2011per, smallwood2016inc}, 
usage of delivery drugs such as pitocin \cite{smallwood2016inc}, 
epidural analgesia \cite{wang2022eff}, 
neonatal hypog \cite{lee2022per}, 
and C-section due to prolonged labor, membrane rupture, or infection (not the C-section itself) \cite{zhang2019ass}. 

Preterm birth increases ASD risk \cite{lord2020aut, crump2021pre}. 
This indicates high CRH, because it is a major trigger of preterm birth \cite{herrera2023pre}, 

\smallest{Direct CRH and ACTH evidence}
Only one paper reported on serum CRH in ASD, and found it higher (4-10yo, n=40) \cite{tsilioni2014ele}. 
With respect to ACTH, the following reports showing increased levels were made. 
Increased serum ACTH and cortisol, with a positive correlation between ACTH and cortisol (6-18yo boys, n=32) \cite{iwata2011inv}. 
Higher plasma ACTH and beta-endorphin, due to more severely affected ASDs (10$\pm$7yo, n=48) \cite{tordjman1997pla}. 
Higher ACTH and lower cortisol (15$\pm$9.5yo, n=36) \cite{curin2003low}. 
Symptoms correlated with ACTH, 16\% with high ACTH, 10\% low basal cortisol, 10\% with decreased cortisol response to ACTH (7.35$\pm$2.6yo, 3-12yo, n=50)  \cite{hamza2010bas}. 
Higher plasma ACTH in Asperger adults, n=20 \cite{tani2005hig}. 

\smallest{Autonomic nervous system}
SNS activity is chronically higher in ASD, as shown by higher basal heart rate, decreased heart rate variability, and blunted skin conductance response \cite{eilam2014abn, neuhaus2016chi, ellenbroek2017aut}. 
Parasympathetic activity is lower, and respiratory sinus arrhythmia data can predict the severity of repetitive behaviors in 5-10yo \cite{condy2017res}. 
Since CRH triggers SNS activation, these data support chronic CRH release.  As noted above, the \thn\ view is that at the time when ASD is diagnosed, CRH release is  diminished to chronically activate CRH1 rather than CRH2. 

\smallest{Mast cells, allergies, gut}
There is high activation of immune mast cells in ASD \cite{theoharides2008nov}, 
and higher prevalence of allergy-related disorders in patients and their siblings \cite{gesundheit2013imm, dai2019inc}. 
CRH directly activated mast cells, which are involved in allergic responses. 

Gut problems are common in ASD patients \cite{wang2011pre, gesundheit2013imm} and their parents \cite{sadik2022par}. 
CRH activation of mast cells increases gut permeability \cite{wallon2008cor}. 

There is general pro-\infl\ immune activation in ASD \cite{gesundheit2013imm}, which can be due to CRH but may simply stem from the excessive release of stress-related agents. 


\smallest{Eye gaze}
CRH overexpression decreases UCN expression in the EW nucleus, which mediates learned eye gaze \cite{kozicz2004uro}. 

\smallest{Other} 
There is high ASD comorbidity with anxiety \cite{reaven2011tre}, and maternal post-traumatic stress disorder increases ASD risk \cite{roberts2014wom}. Both point to high CRH. 

The cortisol response to ACTH stimulation is slower in ASD \cite{marinovic2008slo} and its levels are higher \cite{yang2015cor}, supporting chronic activity. 

ASD is associated with skin hyperpigmentation phenomena \cite{varala2021dia}, 
which are also associated with high ACTH \cite{sandru2019hyp}. 
Recall that melanocortins, which affect skin pigmentation, are a product of the the ACTH gene POMC, stimulated by CRH. 

People with ASD have non-brain health problems in most organs, which are the main cause of premature death \cite{hirvikoski2016pre}. This indirectly supports the UCN path, because UCNs are expressed in these organs and are important for their healthy development. 

\subsection {Anatomical growth}

\smallest{Prenatal}
Intrauterine growth restriction (IUGR) is associated with dramatically increased ASD risk \cite{raghavan2018fet, yalnizoglu2007neu}. 
This supports a growth problem in ASD, and specifically supports high CRH, since high late pregnancy maternal plasma CRH is significantly associated with preterm small for gestational age births \cite{wadhwa2004pla}. 
UCNs are expressed in the placenta and fetal membranes, and are needed for prenatal glucose uptake \cite{imperatore2006uro, simpson2020ucn}. 

\smallest{Early postnatal}
ASD involves increased excitation-inhibition ratio during early  development \cite{rubenstein2003mod}, and decreased lateral inhibition in cortical minicolumns (mean age 12yo) \cite{casanova2015min}. This shows impaired postnatal synaptic pruning and/or impaired postnatal brain development, which is normally dominated by maturation of inhibition.

Accelerated postnatal head circumference growth is consistently reported in ASD \cite{courchesne2001unu, courchesne2003evi, mills2007ele, chawarska2011ear, green2018ske} (but see \cite{dinstein2017no}). 
Baby length was reported to be higher in 4-5 month old boys, with babies at the top 10\% of physical size having greater symptom severity later \cite{chawarska2011ear}. 
Babies with ASD show extremely rapid weight gain in infancy \cite{raghavan2018fet}. 
These data support high postnatal ACTH, since ACTH and other POMC products directly promote bone growth \cite{zhong2005mul}. 

ASD brains show strong local connectivity and low long-distance one \cite{belmonte2004aut}, with hyperconnectivity at young ages \cite{conti2015fir}, supporting reduced postnatal pruning and development. 

Related to sensory hypersensitivity in ASD, there is auditory brainstem pathology, with delayed maturation of low- to high-level pathways \cite{pillion2018aud}. 

Epileptic seizures are common in ASD. This can be explained by increased brain excitation by CRH (e.g., in the \hippo), and by reduced development of brain inhibitory interneurons. 


\subsection {Growth agents \& pathways}
\smallest{Glucose}
ASD is significantly associated with a family history of hypothyroidism \cite{gesundheit2013imm}, especially with maternal hypothyroidism during pregnancy \cite{kaplan2024mat}. 
Hypothyroidism is associated with hypog, further supporting the association of reduced glucose with increased ASD risk. 

Gestational diabetes increases ASD risk \cite{lord2020aut, rowland2021ass}. 
ASD risk is increased with type 1 diabetes and increased blood glucose levels \cite{chen2022type, liu2021poor}. 
Using the HOMA-IR model, people with ASD were shown to have insulin \resist\ and decreased brain glucose \metab\ (4-18yo, n=60) \cite{manco2021cro}. 
These data show the importance of cellular glucose for opposing ASD. 

Neonatal hypog is independently associated with developmental problems and ASD traits \cite{yalnizoglu2007neu}. 
Since UCN secretion is stimulated by glucose, it is decreased in hypog states. 

\smallest{Protein synthesis}
The Akt/mTOR growth axis is dysregulated (mainly reported decreased) in ASD \cite{sheikh2010bdn, tang2014los, nicolini2015dec, russo2015dec, gazestani2019per}. UCNs and GSTRDs are needed for the proper activation of this axis during development. 

Glutathione (GSH), which is a major agent promoting protein synthesis in the PPP, is decreased in ASD \cite{bjorklund2020oxi}. 

ASD is associated with neonatal jaundice \cite{maimburg2008neo, lee2022per} (although not consistently \cite{kujabi2021neo}). 
The enzyme G6PD is essential for the pentose phosphate pathway (PPP) and protein synthesis, and neonatal jaundice indicates G6PD deficiency \cite{lee2022glu} and neonatal hypog \cite{yalnizoglu2007neu}. 
Thus, neonatal jaundice supports reduced cellular glucose in ASD. 

\subsection {Sex steroids}

\smallest {Androgens} DHEA is increased in ASD, in both prepuberty males and females \cite{majewska2014mar, elBaz2014hyp, hassan2019pos, janvsakova2020alt, alZaid2021pot}, showing excessive ASTRDs. 
Note that female androgens are mainly produced in the adrenal gland, where their production is stimulated by ACTH. 

\smallest{\Testos, estrogen} 
Although \testos\ is mainly reported to be increased in ASD (prepubertal, pubertal and in adults) \cite{geier2007pro, schmidtova2010pol, ruta2011inc, alZaid2014alt, elBaz2014hyp, ostatnikova2016neu, hassan2019pos, procyshyn2020eff, muscatello2022sal}, 
it is also reported to be decreased \cite{croonenberghs2010ser, krajmer2011rel, sharpley2017age}
or unchanged \cite{sriram2022eva}. 
The increased DHEA in ASD explains the increased \testos\ results. The conflicting results can be explained by reduced GSTRD production due to high ASTRDs. 

Both an association with higher amniotic fluid estrogens \cite{baron2020foe}
and no association with umbilical cord \testos\ \cite{whitehouse2012per}
were reported. 
No relationship between amniotic fluid \testos\ and autistic traits was found in girls \cite{kung2016traits}. 
There are contradictory reports on the association between prenatal or early postnatal \testos\ and autistic traits \cite{auyeung2012pre, saenz2013pos, kung2016no}. 
Thus, the evidence for a prenatal problem is  weak. 

Follicle-stimulating hormone and estrogen were reported to be decreased in both male and female ASD serum \cite{metwally2020imp}, supporting reduced GSTRDs. 


\smallest{Polycystic ovary syndrome, prenatal androgens} 
PCOS is a condition involving high ASTRDs, mainly DHEA. Large population studies have shown that children of mothers with PCOS have a higher risk of ASD \cite{kosidou2016mat, rotem2021ass}. 
Regardless of a PCOS diagnosis, mothers of children with ASD had higher \testos\ \cite{xu2013mot}. 

As mentioned above, the \thn\ view is that high androgens during sensitive developmental periods increase risk via an association with increased CRH. Prenatal \testos\ is released during weeks 8-20 of pregnancy, so this should be a sensitive period to stress-induced ASD. 
Indeed, the association of ASD with maternal stress during pregnancy \cite{varcin2017pre, seebeck2023ass} may be specific to this period. 
An all-Denmark birth cohort showed that maternal hospitalization due to 1st/2nd trimester viral or bacterial infection increases risk about 3x \cite{atladottir2010mat}, 
and increased 1st trimester maternal stress during the Quebec ice storm was associated with increased ASD risk \cite{walder2014pre}. 

Supporting the effect of stress on sex STRDs, prenatal stress increases anogenital distance (a masculinization index) in human females, with a trend for a decrease in in males \cite{barrett2013pre}. 

Congenital adrenal hyperplasia (CAH) involves high ACTH and androgens, and is not linked to ASD, at least in girls \cite{kung2016traits}. These PCOS and CAH data imply that CRH and ACTH are involved in PCOS but not in CAH pathogenesis.


\smallest{Masculine/feminine features}
ASD involves higher gender dysphoria \cite{glidden2016gen}, 
with males and females more likely to be bisexual and homosexual, respectively \cite{weir2021sex}. 
In the general adult population, people with autistic traits are more likely to be bisexual or with an uncertain orientation \cite{rudolph2018bri}. 
In addition, women with ASD have more masculine facial features than women without ASD \cite{bejerot2012ext, gilani2015sex, tan2017hyp}, 
and men with ASD are more feminine in some ways \cite{bejerot2012ext, gilani2015sex} (although hyper-masculine faces were also reported in prepubertal boys with ASD \cite{tan2017hyp}). 
These data imply lower androgens in males and higher androgens in females during sensitive developmental periods. 
This accords with \thn, because higher CRH-ACTH-DHEA increases androgens in females, and these ASTRDs opposes GSTRDs so decrease the organizational effects of GSTRD androgens in males. 

There is also some evidence that CRH decreases ovary aromatase, which converts \testos\ to estrogen \cite{calogero1996eff}. 
This could be another route through which women with ASD have less estrogen and more \testos\ than other women. 





\smallest{Endocrine disrupting chemicals (EDCs)}
Pre- or perinatal exposure to EDCs (e.g., some pesticides, plastics) increases risk \cite{deCock2012doe, hansen2021pre}. Many EDCs mimic estrogen, amplifying its recruitment of CRH towards the end of pregnancy. (Note that higher postnatal estrogen is relatively protective due to protein synthesis, see above.)

\smallest{Sexual behavior}
CRH1 basally stimulates gonadotropin hormone-releasing hormone (and thus GSTRDs) \cite{phumsatitpong2018est}, which might explain the higher sexual drive and frequent inappropriate sexual behavior shown by adolescents and adults with ASD \cite{schottle2017sex}. 

\smallest{Summary}
ASTRDs are definitely increased in ASD, in both males and females. \Testos\ is usually increased in males, and this is most likely due to the increase in ASTRDs. Conversely, GSTRDs are lower in males. Males are more vulnerable in developmental periods in which they need GSTRDs. These data support increased CRH-ATCH and decreased long-term GSTRD \plast\ in ASD. Prenatal estrogens increase risk, probably by stimulating CRH. 

\subsection {Oxytocin}
OXT is dysregulated in ASD, with most papers reporting decreased OXT \cite{feldman2014par, john2021oxy}, but several reporting an increase \cite{yang2017ser}. 
This accords with the \thn\ prediction. The decrease is due to high \testos\ \cite{dai2017dir} and decreased UCN-mediated \plast\ \cite{bosch2016oxy}. The increase happens because \plast, including at the \hthal, involves maturation of inhibitory interneurons, and possibly because estrogen (which stimulates OXT \cite{sharma2012erb}) is increased due to DHEA. 
%
There is also the possibility of inherently decreased OXT capacity in ASD due to \epig s. 

Pitocin usage during labor indicates delivery problems (see above), but may also increase risk by inducing a long-term \desensit\ of OXT receptors \cite{phaneuf1997des}. 
Usage of the OXT antagonist tractocile during labor is associated with later social communication problems \cite{friedlander2017soc}. 

Several wide cohort studies showed a strong association of ASD with {\bf reduced breastfeeding} \cite{boucher2017ass, tseng2019mat, soke2019ass, chen2021ass, abd2022bre, xiang2023ass, zhan2023inf}. 
This can be because human milk contains OXT \cite{carter2014oxy}, and because breastfeeding is accompanied by longer human touch, which reduces CRH \cite{plotsky1993ear} (even beyond puberty \cite{avishai2001down}) and increases OXT. 

There is one report of decreased fMRI activity in the insula following affective touch in ASD, coupled with increased activity in the primary somatosensory cortex following neutral touch \cite{kaiser2015bra}. 

\subsection {Food intake}
Babies and toddlers later diagnosed with ASD show decreased satiety and increased desire for food intake (non-stop suckling, decreased nursing strike) \cite{lucas2015dys, kara2022nur, peries2023bre, varma2023fee}. 
In the general population, autistic traits at age 6 years are associated with being hungry/not satisfied, or no breastfeeding, or small drinking quantities at age 2 months \cite{vantHof2021ear}. 
These data can be explained by the fact that part of the role of OXT is to signal satiety and decrease food intake (mainly sugars) \cite{klockars2015cen}, so reduced OXT prolongs the time to satiety. 
Another account is that the deficiency of glucose for growth in ASD is sensed (via an unknown mechanism, although see CRRs below), and this stimulates continued food intake. 

Children with ASD show restricted food preference and an unwillingness to try novel food \cite{wallace2018aut}. 
This may be due to their general avoidance of novelty to prevent higher CRH release. 

At later ages, there is high obesity in ASD \cite{croen2015hea}, 
which could be due to chronically high activation of CRH1, which stimulates the counter-regulatory responses (CRRs) to hypog and food intake \cite{mccrimmon2006cor}. 
In addition, in ASD, less ingested glucose is channelled to protein synthesis, which leaves excessive blood glucose. In this situation, serotonin promotes the storage of glucose in adipose tissue as fat \cite{yabut2019eme}. 
Supporting this account, blood serotonin is elevated in patients \cite{muller2016ser} and correlates with symptom severity \cite{lin2024ris}. 


\section{Treatment}
\subsection{Detection}
Postnatal levels of blood, urine, saliva, hari and/or CSF CRH, ACTH, cortisol, UCNs, \testos, estrogen, DHEA/cortisol ratio, and OXT, can be used for early screening. The most promising biomarker may be {\bf ACTH}, because it is directly related to both CRH and ASTRDs. 
Levels at birth would probably be less informative than levels at the age of 3 months, since at this age adrenal involution, coupled with a dramatic decrease in DHEA, should already have happened. However, as indicated above, there are reports in which adrenal involution occurs much earlier. Normal timelines for CRH and ACTH decrease and adrenal involution should be investigated more in order to devise the optimal detection schedule for ASD. 
Prenatal (maternal plasma and amniotic fluid) levels, mainly of CRH and ACTH, would  be useful too. 

Direct demonstration of higher adrenal volume using imaging would also point to a problem, because reduced ACTH is a major factor in adrenal involution \cite{lazo1960pit}, with both CRH and ACTH increasing adrenal volume \cite{miskowiak1986com}. 

\subsection{Preventive treatment}
\smallestq{Is it justified}
When discussing preventive treatment, we must keep in mind that  so-called `autistic traits' are part of the vast repertoire of human properties, and are not necessarily a bad thing to have. Nonetheless, it is clear that many people with ASD are simply unable to conduct an independent fulfilling life, which justifies treatment. 

\smallest{Definition}
By `preventive treatment', I mean treatment made during the damage period, since it is generally not possible to reduce high pre- and perinatal CRH. Nonetheless, it may be a good idea not to take certain medications during pregnancy. For example, selective serotonin reuptake inhibitors, which are a very common treatment for depression, are associated with increased 2nd trimester CRH (although causality has not been shown) \cite{hannerfors2015tre}. 

\smallest{Behavioral}
 A general principle is that drug treatment should be accompanied by behavioral treatment. The problem is impaired brain \plast\ (learning), but the brain still has learning capacity, and the impairment can be greatly alleviated by repeated training \cite{sandbank2020pro}. 

 Pleasant postnatal human {\bf touch}, and {\bf breastfeeding}, should reduce risk by increasing OXT and reducing CRH in the brain. 

\smallest{CRH, DHT}
 The best treatment would be to {\bf reduce CRH levels}. Antibodies to CRH have been isolated in mice \cite{futch2019ant}. However, the main damage is done in the brain. RNA interference methods can be used to reduce CRH production, but the drug needs to penetrate the brain. 
 
There is a promising way to reduce CRH. DHT mediates many of the organizational effects of GSTRD androgens \cite{biason2013mar}.
Probably as part of this role, it suppresses ASTRDs, decreasing HPA axis activity and the release of its driving agents, including CRH and ACTH \cite{lund2006and, handa2013cen}. 
DHT synthesis during male minipuberty takes place only in the testis, with adrenal production being extremely high at birth but dropping to very low levels at 10 weeks \cite{dhayat2017and}. 
Conversely, ACTH inhibits 5a-reductase, the rate limiting enzyme in DHT production \cite{kitay1971eff}. 

Thus, {\bf exogenous DHT, which is readily available, could ameliorate or even prevent the ASD phenotype} if used at the right time. DHT is available in transdermal patch form, making its administration to babies very easy. 

Obviously, this treatment may be suitable only for males, but males comprise 80\% of the ASD population. 
DHT mediates some or all of its anti-CRH effects via its conversion to 3b-diol (by 3beta-hydroxysteroid dehydrogenase) \cite{lund2006and}, which binds estrogen receptor beta but not the androgen receptor. Like estrogen, 3b-diol phasically increases OXT \cite{sharma2012erb}. Thus, 3b-diol treatment may be suitable for females (and possibly males as well). 



This DHT proposal may seem paradoxical in light of the increased levels of androgens reported in ASD. However, recall that according to \thn, these increased levels are mainly due to ASTRDs, not GSTRDs, which are the normal source of DHT during minipuberty. 

\smallest{Other}
There are additional possible treatments. Cortisol reduces CRH release, so postnatal {\bf exogenous glucocorticoids (GCs)} might help. GC doses should be lower than those that desensitize the GC receptor (GR). 

 {\bf ACTH suppression} (e.g., with somatostatin analogues such as pasireotide) could be helpful, mainly to reduce damage done in the ASTRD. This treatment may need to be supplemented by exogenous GCs. 

 {\bf Exogenous OXT} during the first postnatal months might be beneficial. Some intranasal OXT crosses the blood-brain barrier (BBB) \cite{lee2020lab}, but there is no evidence that it reaches the hypothalamus. Blood OXT needs assistance in order to cross the BBB in meaningful amounts and to remain active for more than a short time. 

 Exogenous UCNs diffuse across the blood-brain barrier \cite{pan2008uro}, and may be beneficial by competing with CRH for receptor binding.  



\subsection{Symptomatic treatment}
ACTH suppression (e.g., with DHT) may be a good treatment for aggression symptoms exhibited after the initial damage period (recall that ACTH and DHEA promote such aggression). Reducing CRH should also counter aversion to novelty and repetitive movements and habits. 

\section{Discussion}
ASD is one of the major brain disorders, and one of the largest mysteries of modern science, medicine and human society. Here I presented the first complete theory of ASD, complete in the sense that it explains the etiology, symptoms, and pathology of ASD, and relies on substantial empirical evidence. 

\smallest{Other theories}
At present, there is no theory of ASD that mechanistically explains even its symptoms, let alone its etiology and pathology \cite{yenkoyan2017adv}. 
Several hypotheses have been articulated regarding the nature of ASD. The `extreme male brain' hypothesis \cite{baron2010emp}
posits a difference between systematizing, which is supposedly a male way of behavior, and empathizing, supposedly female. These notions are very hard to define, and in any case do not provide any mechanistic account. \thn\ agrees that females with ASD are more masculine in some ways, but as the evidence brought above shows, males with ASD can be argued to be more feminine than males without ASD. 

The idea that OXT is central in ASD has been raised \cite{carter2007sex}, which is natural in light of the role of OXT is social development. 
However, mechanisms have not been detailed, and although OXT dysregulation exists in ASD, it is clear that OXT is far from explaining the variety of ASD symptoms and properties. 
In \thn, the main downstream event is an impairment in postnatal \plast\ due to reduced growth signaling by UCNs and GSTRDs. OXT is one of the main sites of damage. 

Another idea is that ASD involves an increased excitation-inhibition (E/I)  ratio during early development \cite{rubenstein2003mod}. This is indeed a plausible outcome of impaired postnatal \plast, but a detailed review concluded that although excitation and inhibition are altered in ASD, there is no real evidence for the increased E/I theory \cite{dickinson2016mea}. 

A hypothesis with some mechanistic basis is that of aberrant growth pathways such as Wnt, ERK, and Akt \cite{courchesne2020pre}. In this idea, ASD begins in prenatal life, with a disruption of proliferation and migration. This follows with a disrupted postnatal brain connectivity. There is an emphasis on risk genes. 
\thn\ agrees with this hypothesis in the importance of impaired growth pathways, but disagrees with it in their timing and concrete identity, and in the genetic emphasis. In addition, this theory does not include any mechanistic account of the specific symptoms of ASD. 

\smallest{Strengths and weaknesses}
\thn\ presents a view of the etiology of ASD that is highly biologically plausible and consistent with the evidence. It mechanistically addresses basically all of the symptoms of ASD, and utilizes all of the salient data gathered about the disorder. It is supported by a substantial body of empirical evidence of different types and modalities. As such, it is a very strong theory. Its greatest weakness at present is that there is no supporting evidence showing that the putative pathways are indeed impaired during the relevant damage period. There is a lot of evidence for the involvement of CRH and sex steroids somewhat after (and before) this period, but not during the first 1-9 postnatal months. In addition, there is very little UCN data. 

\smallest{Genetics} 
Because ASD occurs due to stress during pregnancy or after birth, \thn\ does not predict meaningful DNA changes. The tendency for increased CRH release (and/or for increased effect on its downstream paths) may be encoded genetically or epigenetically, but this most likely involves a large number of genes and a small effect in each of them. The damage occurs during development, and may not leave any genetic or epigenetic trace in the causal genes, although there is likely a persistent epigenetic change to the CRH system that causes chronic CRH1 activation. 
Indeed, there are different DNA methylation patterns between disease-discordant monozygotic twins, with a significant correlation between DNA methylation and ASD traits \cite{wong2014met}. 
Overall, \thn\ generally does not expect supporting genetic evidence. There is some hypothesized genetic data pointing to CRH, CRH2 and UCN3 genes \cite{frasch2023aut}, but it is not strong. 

\smallest{Treatment} 
An exciting aspect of \thn\ is that it points to easy biomarkers (ACTH, ASTRDs, adrenal volume) and to a readily available preventive treatment that's quite easy to administer to babies. Although the definition of what constitutes `excessive' CRH/ACTH would surely be blurry, many of the more severe cases of ASD most likely involve levels that would be classified as excessive without argument. The proposed treatment may alleviate these cases, while people with ASD traits will not be treated, allowing human diversity. 

\section*{List of Abbreviations}
Terms introduced in this paper are marked by a {\bf bold} font. 

ACTH: \acth. 

aMSH: alpha-melanocyte stimulating hormone. 

ASD : \asd.

{\bf ASTRD: adrenal steroid.}

BNST: \bnst. 

{\bf C-CRH2: CRH-induced CRH2 receptor signaling.}

cAMP: cyclic adenosine monophosphate. 

CRH: \crh\ (also CRF, factor). 

DHT: \dht. 


EW: Edinger-Westphal (nucleus). 

{\bf GSTRD: gonadal steroid.}

DHEA(S): \dhea\ (sulfated). 

GC: glucocorticoid. 

MAPK: mitogen-activated protein kinase. 

OXT: \oxt. 

PCOS: \pcos. 

POMC: prooptiomelanocortin. 

PPP: pentose phosphate pathway. 

SNS: \sympa\ nervous system. 

STRD: steroid. 

{\bf \thn: the theory of ASD presented here.}

{\bf U-CRH2: UCN-induced CRH2 receptor signaling.}

\bibliographystyle{vancouver} 
\bibliography{ASD-anrx-etc,mdd,msc,adhd} 

\begin{thebibliography}{100}

\bibitem{lord2020aut}
Lord C, Brugha TS, Charman T, Cusack J, Dumas G, Frazier T, et~al.
\newblock Autism spectrum disorder.
\newblock Nature reviews Disease primers. 2020;6(1):1-23.

\bibitem{hirota2023aut}
Hirota T, King BH.
\newblock Autism spectrum disorder: A review.
\newblock Jama. 2023;329(2):157-68.

\bibitem{maenner2023pre}
Maenner MJ.
\newblock Prevalence and characteristics of autism spectrum disorder among children aged 8 years—Autism and Developmental Disabilities Monitoring Network, 11 sites, {U}nited {S}tates, 2020.
\newblock MMWR Surveillance Summaries. 2023;72.

\bibitem{bale2004crf}
Bale TL, Vale WW.
\newblock {CRF} and {CRF} receptors: role in stress responsivity and other behaviors.
\newblock Annu Rev Pharmacol Toxicol. 2004;44:525-57.

\bibitem{deussing2018cor}
Deussing JM, Chen A.
\newblock The corticotropin-releasing factor family: physiology of the stress response.
\newblock Physiological reviews. 2018;98(4):2225-86.

\bibitem{li2003uro}
Li C, Chen P, Vaughan J, Blount A, Chen A, Jamieson PM, et~al.
\newblock Urocortin {III} is expressed in pancreatic $\beta$-cells and stimulates insulin and glucagon secretion.
\newblock Endocrinology. 2003;144(7):3216-24.

\bibitem{li2007uro}
Li C, Chen P, Vaughan J, Lee KF, Vale W.
\newblock Urocortin 3 regulates glucose-stimulated insulin secretion and energy homeostasis.
\newblock Proceedings of the National Academy of Sciences. 2007;104(10):4206-11.

\bibitem{simpson2020ucn}
Simpson SJ, Smith LI, Jones PM, Bowe JE.
\newblock {UCN}2: a new candidate influencing pancreatic $\beta$-cell adaptations in pregnancy.
\newblock Journal of Endocrinology. 2020;245(2):247-57.

\bibitem{chanalaris2005hyp}
Chanalaris A, Lawrence KM, Townsend PA, Davidson S, Jashmidi Y, Stephanou A, et~al.
\newblock Hypertrophic effects of urocortin homologous peptides are mediated via activation of the {A}kt pathway.
\newblock Biochemical and biophysical research communications. 2005;328(2):442-8.

\bibitem{markovic2011mol}
Markovic D, Punn A, Lehnert H, Grammatopoulos DK.
\newblock Molecular determinants and feedback circuits regulating type 2 {CRH} receptor signal integration.
\newblock Biochimica et Biophysica Acta (BBA)-Molecular Cell Research. 2011;1813(5):896-907.

\bibitem{huang2012his}
Huang HY, Liu DD, Chang HF, Chen WF, Hsu HR, Kuo JS, et~al.
\newblock Histone deacetylase inhibition mediates urocortin-induced antiproliferation and neuronal differentiation in neural stem cells.
\newblock Stem Cells. 2012;30(12):2760-73.

\bibitem{roustit2014uro}
Roustit MM, Vaughan JM, Jamieson PM, Cleasby ME.
\newblock Urocortin 3 activates {AMPK} and {AKT} pathways and enhances glucose disposal in rat skeletal muscle.
\newblock The Journal of endocrinology. 2014;223(2):143.

\bibitem{walther2014uro}
Walther S, Pluteanu F, Renz S, Nikonova Y, Maxwell JT, Yang LZ, et~al.
\newblock Urocortin 2 stimulates nitric oxide production in ventricular myocytes via {A}kt-and {PKA}-mediated phosphorylation of e{NOS} at serine 1177.
\newblock American Journal of Physiology-Heart and Circulatory Physiology. 2014;307(5):H689-700.

\bibitem{gao2016one}
Gao MH, Giamouridis D, Lai NC, Walenta E, Paschoal VA, Kim YC, et~al.
\newblock One-time injection of {AAV8} encoding urocortin 2 provides long-term resolution of insulin resistance.
\newblock JCI insight. 2016;1(15).

\bibitem{kavalakatt2021uro}
Kavalakatt S, Khadir A, Madhu D, Koistinen HA, Al-Mulla F, Tuomilehto J, et~al.
\newblock Urocortin 3 overexpression reduces {ER} stress and heat shock response in {3T3-L1} adipocytes.
\newblock Scientific Reports. 2021;11(1):15666.

\bibitem{meister2022viv}
Meister J, Bone DB, Knudsen JR, Barella LF, Liu L, Lee R, et~al.
\newblock In vivo metabolic effects after acute activation of skeletal muscle {G}s signaling.
\newblock Molecular Metabolism. 2022;55:101415.

\bibitem{lautherbach2022uro}
Lautherbach N, Gon{\c{c}}alves DA, Silveira WA, Paula-Gomes S, Valentim RR, Zanon NM, et~al.
\newblock Urocortin 2 promotes hypertrophy and enhances skeletal muscle function through c{AMP} and insulin/{IGF}-1 signaling pathways.
\newblock Molecular Metabolism. 2022;60:101492.

\bibitem{flaherty2023chr}
Flaherty~III SE, Bezy O, Zheng W, Yan D, Li X, Jagarlapudi S, et~al.
\newblock Chronic {UCN2} treatment desensitizes {CRHR2} and improves insulin sensitivity.
\newblock Nature Communications. 2023;14(1):3953.

\bibitem{mccrimmon2006cor}
McCrimmon RJ, Song Z, Cheng H, McNay EC, Weikart-Yeckel C, Fan X, et~al.
\newblock Corticotrophin-releasing factor receptors within the ventromedial hypothalamus regulate hypoglycemia-induced hormonal counterregulation.
\newblock The Journal of clinical investigation. 2006;116(6):1723-30.

\bibitem{hauger2013des}
Hauger RL, Olivares-Reyes JA, Braun S, Hernandez-Aranda J, Hudson CC, Gutknecht E, et~al.
\newblock Desensitization of human {CRF2} (a) receptor signaling governed by agonist potency and $\beta$arrestin2 recruitment.
\newblock Regulatory peptides. 2013;186:62-76.

\bibitem{croen2015hea}
Croen LA, Zerbo O, Qian Y, Massolo ML, Rich S, Sidney S, et~al.
\newblock The health status of adults on the autism spectrum.
\newblock Autism. 2015;19(7):814-23.

\bibitem{turcu2014adr}
Turcu A, Smith JM, Auchus R, Rainey WE.
\newblock Adrenal androgens and androgen precursors: definition, synthesis, regulation and physiologic actions.
\newblock Comprehensive Physiology. 2014;4(4):1369.

\bibitem{kageyama2013reg}
Kageyama K.
\newblock Regulation of gonadotropins by corticotropin-releasing factor and urocortin.
\newblock Frontiers in endocrinology. 2013;4:12.

\bibitem{norman2004ste}
Norman AW, Mizwicki MT, Norman DP.
\newblock Steroid-hormone rapid actions, membrane receptors and a conformational ensemble model.
\newblock Nat Revs Drug Discovery. 2004;3(1):27-41.

\bibitem{clark2018mec}
Clark BJ, Prough RA, Klinge CM.
\newblock Mechanisms of action of dehydroepiandrosterone.
\newblock Vitamins and hormones. 2018;108:29-73.

\bibitem{saponaro2007lon}
Saponaro S, Guarnieri V, Pescarmona GP, Silvagno F.
\newblock Long-term exposure to dehydroepiandrosterone affects the transcriptional activity of the glucocorticoid receptor.
\newblock The Journal of steroid biochemistry and molecular biology. 2007;103(2):129-36.

\bibitem{mcnelis2013deh}
McNelis JC, Manolopoulos KN, Gathercole LL, Bujalska IJ, Stewart PM, Tomlinson JW, et~al.
\newblock Dehydroepiandrosterone exerts antiglucocorticoid action on human preadipocyte proliferation, differentiation, and glucose uptake.
\newblock American Journal of Physiology-Endocrinology and Metabolism. 2013;305(9):E1134-44.

\bibitem{ben2007par}
Ben-David S, Zuckerman-Levin N, Epelman M, Shen-Orr Z, Levin M, Sujov P, et~al.
\newblock Parturition itself is the basis for fetal adrenal involution.
\newblock The Journal of Clinical Endocrinology \& Metabolism. 2007;92(1):93-7.

\bibitem{dhayat2017and}
Dhayat NA, Dick B, Frey BM, d’Uscio CH, Vogt B, Fl{\"u}ck CE.
\newblock Androgen biosynthesis during minipuberty favors the backdoor pathway over the classic pathway: Insights into enzyme activities and steroid fluxes in healthy infants during the first year of life from the urinary steroid metabolome.
\newblock The Journal of steroid biochemistry and molecular biology. 2017;165:312-22.

\bibitem{carter2014oxy}
Carter CS.
\newblock Oxytocin pathways and the evolution of human behavior.
\newblock Annu Rev Psychol. 2014;65:17-39.

\bibitem{frankiensztajn2018com}
Frankiensztajn LM, Gur-Pollack R, Wagner S.
\newblock A combinatorial modulation of synaptic plasticity in the rat medial amygdala by oxytocin, urocortin3 and estrogen.
\newblock Psychoneuroendocrinology. 2018;92:95-102.

\bibitem{ohno2018inc}
Ohno S, Hashimoto H, Fujihara H, Fujiki N, Yoshimura M, Maruyama T, et~al.
\newblock Increased oxytocin-monomeric red fluorescent protein 1 fluorescent intensity with urocortin-like immunoreactivity in the hypothalamo-neurohypophysial system of aged transgenic rats.
\newblock Neuroscience Research. 2018;128:40-9.

\bibitem{dai2017dir}
Dai D, Li QC, Zhu QB, Hu SH, Balesar R, Swaab D, et~al.
\newblock Direct Involvement of Androgen Receptor in Oxytocin Gene Expression: Possible Relevance for Mood Disorders.
\newblock Neuropsychopharmacology. 2017.

\bibitem{chu2013eff}
Chu CP, Jin WZ, Bing YH, Jin QH, Kannan H, Qiu DL.
\newblock Effects of stresscopin on rat hypothalamic paraventricular nucleus neurons in vitro.
\newblock PloS one. 2013;8(1):e53863.

\bibitem{bosch2016oxy}
Bosch OJ, Dabrowska J, Modi ME, Johnson ZV, Keebaugh AC, Barrett CE, et~al.
\newblock Oxytocin in the nucleus accumbens shell reverses {CRFR}2-evoked passive stress-coping after partner loss in monogamous male prairie voles.
\newblock Psychoneuroendocrinology. 2016;64:66-78.

\bibitem{martinon2018cor}
Martinon D, Dabrowska J.
\newblock Corticotropin-releasing factor receptors modulate oxytocin release in the dorsolateral bed nucleus of the stria terminalis ({BNST}) in male rats.
\newblock Frontiers in neuroscience. 2018;12:337055.

\bibitem{matsuzaki1989eff}
Takamatsu Y, Moroji T.
\newblock The effects of intracerebroventricularly injected corticotropin-releasing factor ({CRF}) on the central nervous system: behavioural and biochemical studies.
\newblock Neuropeptides. 1989;13(3):147-55.

\bibitem{strome2002int}
Strome EM, Wheler GT, Higley JD, Loriaux DL, Suomi SJ, Doudet DJ.
\newblock Intracerebroventricular corticotropin-releasing factor increases limbic glucose metabolism and has social context-dependent behavioral effects in nonhuman primates.
\newblock Proceedings of the National Academy of Sciences. 2002;99(24):15749-54.

\bibitem{bakshi2007sti}
Bakshi VP, Newman SM, Smith-Roe S, Jochman KA, Kalin NH.
\newblock Stimulation of lateral septum {CRF2} receptors promotes anorexia and stress-like behaviors: functional homology to {CRF1} receptors in basolateral amygdala.
\newblock Journal of neuroscience. 2007;27(39):10568-77.

\bibitem{carpenter2007cor}
Carpenter RE, Watt MJ, Forster GL, {\O}verli {\O}, Bockholt C, Renner KJ, et~al.
\newblock Corticotropin releasing factor induces anxiogenic locomotion in trout and alters serotonergic and dopaminergic activity.
\newblock Hormones and behavior. 2007;52(5):600-11.

\bibitem{ogino2014com}
Ogino M, Okumura A, Khan MSI, Cline MA, Tachibana T.
\newblock Comparison of brain urocortin-3 and corticotrophin-releasing factor for physiological responses in chicks.
\newblock Physiology \& behavior. 2014;125:57-61.

\bibitem{vanErp1993dif}
Van~Erp A, Kruk M, Van~Oers H, Hemmers N.
\newblock Differential effect of {ACTH}1--24 and $\alpha$-{MSH} induced grooming in the paraventricular nucleus of the hypothalamus.
\newblock Brain research. 1993;603(2):296-301.

\bibitem{wikberg2000new}
Wikberg JE, Muceniece R, Mandrika I, Prusis P, Lindblom J, Post C, et~al.
\newblock New aspects on the melanocortins and their receptors.
\newblock Pharmacological research. 2000;42(5):393-420.

\bibitem{sherman1986icv}
Sherman JE, Kalin NH.
\newblock {ICV-CRH} potently affects behavior without altering antinociceptive responding.
\newblock Life sciences. 1986;39(5):433-41.

\bibitem{gesing2001psy}
Gesing A, Bilang-Bleuel A, Droste SK, Linthorst AC, Holsboer F, Reul JM.
\newblock Psychological stress increases hippocampal mineralocorticoid receptor levels: involvement of corticotropin-releasing hormone.
\newblock Journal of Neuroscience. 2001;21(13):4822-9.

\bibitem{roussel2005gen}
Montigny D, Hemsworth P.
\newblock Gender-specific effects of prenatal stress on emotional reactivity and stress physiology of goat kids.
\newblock Hormones and behavior. 2005;47(3):256-66.

\bibitem{shemesh2016ucn}
Shemesh Y, Forkosh O, Mahn M, Anpilov S, Sztainberg Y, Manashirov S, et~al.
\newblock Ucn3 and {CRF-R}2 in the medial amygdala regulate complex social dynamics.
\newblock Nature neuroscience. 2016;19(11):1489-96.

\bibitem{takesian2013bal}
Takesian AE, Hensch TK.
\newblock Balancing plasticity/stability across brain development.
\newblock Progress in brain research. 2013;207:3-34.

\bibitem{quast2017dev}
Quast KB, Ung K, Froudarakis E, Huang L, Herman I, Addison AP, et~al.
\newblock Developmental broadening of inhibitory sensory maps.
\newblock Nature Neuroscience. 2017;20(2):189-99.

\bibitem{steenfeldt2020pre}
Steenfeldt-Kristensen C, Jones CA, Richards C.
\newblock The prevalence of self-injurious behaviour in autism: a meta-analytic study.
\newblock Journal of autism and developmental disorders. 2020;50:3857-73.

\bibitem{yamamoto2010inv}
Yamamoto A, Sugimoto Y.
\newblock Involvement of peripheral mu opioid receptors in scratching behavior in mice.
\newblock European journal of pharmacology. 2010;649(1-3):336-41.

\bibitem{fetissov2006agg}
Fetissov SO, Hallman J, Nilsson I, Lefvert AK, Oreland L, H{\"o}kfelt T.
\newblock Aggressive behavior linked to corticotropin-reactive autoantibodies.
\newblock Biological psychiatry. 2006;60(8):799-802.

\bibitem{veenema2007low}
Veenema AH, Torner L, Blume A, Beiderbeck DI, Neumann ID.
\newblock Low inborn anxiety correlates with high intermale aggression: link to {ACTH} response and neuronal activation of the hypothalamic paraventricular nucleus.
\newblock Hormones and behavior. 2007;51(1):11-9.

\bibitem{pajer2006adr}
Pajer K, Tabbah R, Gardner W, Rubin RT, Czambel RK, Wang Y.
\newblock Adrenal androgen and gonadal hormone levels in adolescent girls with conduct disorder.
\newblock Psychoneuroendocrinology. 2006;31(10):1245-56.

\bibitem{soma2015dhe}
Soma KK, Rendon NM, Boonstra R, Albers HE, Demas GE.
\newblock {DHEA} effects on brain and behavior: insights from comparative studies of aggression.
\newblock The Journal of steroid biochemistry and molecular biology. 2015;145:261-72.

\bibitem{mcglone2014dis}
McGlone F, Wessberg J, Olausson H.
\newblock Discriminative and affective touch: sensing and feeling.
\newblock Neuron. 2014;82(4):737-55.

\bibitem{kozicz2011edi}
Kozicz T, Bittencourt JC, May PJ, Reiner A, Gamlin PD, Palkovits M, et~al.
\newblock The {E}dinger-{W}estphal nucleus: A historical, structural, and functional perspective on a dichotomous terminology.
\newblock Journal of Comparative Neurology. 2011;519(8):1413-34.

\bibitem{cservenka2010pos}
Cservenka A, Spangler E, Cote DM, Ryabinin AE.
\newblock Postnatal developmental profile of urocortin 1 and cocaine-and amphetamine-regulated transcript in the perioculomotor region of {C57BL/6J} mice.
\newblock Brain research. 2010;1319:33-43.

\bibitem{lim2021ear}
Lim YH, Licari M, Spittle AJ, Watkins RE, Zwicker JG, Downs J, et~al.
\newblock Early motor function of children with autism spectrum disorder: a systematic review.
\newblock Pediatrics. 2021;147(2).

\bibitem{samuelsson2004cor}
Samuelsson S, Lange JS, Hinkle RT, Tarnopolsky M, Isfort RJ.
\newblock Corticotropin-releasing factor 2 receptor localization in skeletal muscle.
\newblock Journal of Histochemistry \& Cytochemistry. 2004;52(7):967-77.

\bibitem{hinkle2003act}
Hinkle RT, Donnelly E, Cody DB, Samuelsson S, Lange JS, Bauer MB, et~al.
\newblock Activation of the {CRF} 2 receptor modulates skeletal muscle mass under physiological and pathological conditions.
\newblock American Journal of Physiology-Endocrinology and Metabolism. 2003;285(4):E889-98.

\bibitem{jamieson2011uro}
Jamieson P, Cleasby M, Morton N, Kelly P, Brownstein D, Mustard K, et~al.
\newblock Urocortin 3 transgenic mice exhibit a metabolically favourable phenotype resisting obesity and hyperglycaemia on a high-fat diet.
\newblock Diabetologia. 2011;54:2392-403.

\bibitem{reutenauer2012uro}
Reutenauer-Patte J, Boittin FX, Patthey-Vuadens O, Ruegg UT, Dorchies OM.
\newblock Urocortins improve dystrophic skeletal muscle structure and function through both {PKA}-and {E}pac-dependent pathways.
\newblock The American journal of pathology. 2012;180(2):749-62.

\bibitem{azcoitia2019mol}
Azcoitia I, Barreto GE, Garcia-Segura LM.
\newblock Molecular mechanisms and cellular events involved in the neuroprotective actions of estradiol. Analysis of sex differences.
\newblock Frontiers in neuroendocrinology. 2019;55:100787.

\bibitem{kuiri2014act}
Kuiri-H{\"a}nninen T, Sankilampi U, Dunkel L.
\newblock Activation of the hypothalamic-pituitary-gonadal axis in infancy: minipuberty.
\newblock Hormone research in paediatrics. 2014;82(2):73-80.

\bibitem{derks2010sex}
Derks NM, Gaszner B, Roubos EW, Kozicz LT.
\newblock Sex differences in urocortin 1 dynamics in the non-preganglionic {E}dinger--{W}estphal nucleus of the rat.
\newblock Neuroscience research. 2010;66(1):117-23.

\bibitem{sharma2012erb}
Sharma D, Handa RJ, Uht RM.
\newblock The {ER}$\beta$ ligand 5$\alpha$-androstane, 3$\beta$, 17$\beta$-diol (3$\beta$-diol) regulates hypothalamic oxytocin ({O}xt) gene expression.
\newblock Endocrinology. 2012;153(5):2353-61.

\bibitem{hiroi2013and}
Hiroi R, Lacagnina AF, Hinds LR, Carbone DG, Uht RM, Handa RJ.
\newblock The androgen metabolite, 5$\alpha$-androstane-3$\beta$, 17$\beta$-diol (3$\beta$-diol), activates the oxytocin promoter through an estrogen receptor-$\beta$ pathway.
\newblock Endocrinology. 2013;154(5):1802-12.

\bibitem{zwaigenbaum2005beh}
Zwaigenbaum L, Bryson S, Rogers T, Roberts W, Brian J, Szatmari P.
\newblock Behavioral manifestations of autism in the first year of life.
\newblock International journal of developmental neuroscience. 2005;23(2-3):143-52.

\bibitem{hu2020ear}
Hu P, Maita I, Phan ML, Gu E, Kwok C, Dieterich A, et~al.
\newblock Early-life stress alters affective behaviors in adult mice through persistent activation of {CRH-BDNF} signaling in the oval bed nucleus of the stria terminalis.
\newblock Translational psychiatry. 2020;10(1):396.

\bibitem{howlin2009sav}
Howlin P, Goode S, Hutton J, Rutter M.
\newblock Savant skills in autism: psychometric approaches and parental reports.
\newblock Philosophical Transactions of the Royal Society B: Biological Sciences. 2009;364(1522):1359-67.

\bibitem{meilleur2015pre}
Meilleur AAS, Jelenic P, Mottron L.
\newblock Prevalence of clinically and empirically defined talents and strengths in autism.
\newblock Journal of autism and developmental disorders. 2015;45:1354-67.

\bibitem{gardner2020ass}
Gardner RM, Dalman C, Rai D, Lee BK, Karlsson H.
\newblock The association of paternal {IQ} with autism spectrum disorders and its comorbidities: a population-based cohort study.
\newblock Journal of the American Academy of Child \& Adolescent Psychiatry. 2020;59(3):410-21.

\bibitem{roozendaal2011mem}
Roozendaal B, McGaugh JL.
\newblock Memory modulation.
\newblock Behavioral neuroscience. 2011;125(6):797.

\bibitem{seale2004gon}
Seale J, Wood S, Atkinson H, Bate E, Lightman S, Ingram C, et~al.
\newblock Gonadectomy reverses the sexually diergic patterns of circadian and stress-induced hypothalamic-pituitary-adrenal axis activity in male and female rats.
\newblock Journal of neuroendocrinology. 2004;16(6):516-24.

\bibitem{gardener2011per}
Gardener H, Spiegelman D, Buka SL.
\newblock Perinatal and neonatal risk factors for autism: a comprehensive meta-analysis.
\newblock Pediatrics. 2011;128(2):344-55.

\bibitem{smallwood2016inc}
Smallwood M, Sareen A, Baker E, Hannusch R, Kwessi E, Williams T.
\newblock Increased risk of autism development in children whose mothers experienced birth complications or received labor and delivery drugs.
\newblock ASN neuro. 2016;8(4):1759091416659742.

\bibitem{wang2022eff}
Wang X, Li J, Liu D.
\newblock Effects of epidural analgesia exposure during parturition on autism spectrum disorder in newborns: A systematic review and meta-analysis based on cohort study.
\newblock Frontiers in psychiatry. 2022;13:974596.

\bibitem{lee2022per}
Lee IC, Wang YH, Chiou JY, Wei JCC.
\newblock Perinatal factors in newborn are insidious risk factors for childhood autism spectrum disorders: a population-based study.
\newblock Journal of Autism and Developmental Disorders. 2022;52(1):52-60.

\bibitem{zhang2019ass}
Zhang T, Sidorchuk A, Sevilla-Cerme{\~n}o L, Vilaplana-P{\'e}rez A, Chang Z, Larsson H, et~al.
\newblock Association of cesarean delivery with risk of neurodevelopmental and psychiatric disorders in the offspring: a systematic review and meta-analysis.
\newblock JAMA network open. 2019;2(8):e1910236-6.

\bibitem{crump2021pre}
Crump C, Sundquist J, Sundquist K.
\newblock Preterm or early term birth and risk of autism.
\newblock Pediatrics. 2021;148(3).

\bibitem{herrera2023pre}
Herrera CL, Maiti K, Smith R.
\newblock Preterm birth and corticotrophin-releasing hormone as a placental clock.
\newblock Endocrinology. 2023;164(2):bqac206.

\bibitem{tsilioni2014ele}
Tsilioni I, Dodman N, Petra AI, Taliou A, Francis K, Moon-Fanelli A, et~al.
\newblock Elevated serum neurotensin and {CRH} levels in children with autistic spectrum disorders and tail-chasing {B}ull {T}erriers with a phenotype similar to autism.
\newblock Translational Psychiatry. 2014;4(10):e466-6.

\bibitem{iwata2011inv}
Iwata K, Matsuzaki H, Miyachi T, Shimmura C, Suda S, Tsuchiya KJ, et~al.
\newblock Investigation of the serum levels of anterior pituitary hormones in male children with autism.
\newblock Molecular autism. 2011;2:1-6.

\bibitem{tordjman1997pla}
Tordjman S, McBride PA, Hertzig ME, Snow ME, Anderson GM, Hall LM, et~al.
\newblock Plasma $\beta$-endorphin, adrenocorticotropin hormone, and cortisol in autism.
\newblock Journal of Child Psychology and Psychiatry. 1997;38(6):705-15.

\bibitem{curin2003low}
{\'C}urin JM, Terzi{\'c} J, Petkovi{\'c} ZB, Zekan L, Terzi{\'c} IM, {\v{S}}u{\v{s}}njara IM.
\newblock Lower cortisol and higher {ACTH} levels in individuals with autism.
\newblock Journal of autism and developmental disorders. 2003;33:443-8.

\bibitem{hamza2010bas}
Hamza RT, Hewedi DH, Ismail MA.
\newblock Basal and adrenocorticotropic hormone stimulated plasma cortisol levels among {E}gyptian autistic children: relation to disease severity.
\newblock Italian Journal of Pediatrics. 2010;36:1-6.

\bibitem{tani2005hig}
Tani P, Lindberg N, Matto V, Appelberg B, Nieminen-von Wendt T, von Wendt L, et~al.
\newblock Higher plasma {ACTH} levels in adults with {A}sperger syndrome.
\newblock Journal of Psychosomatic Research. 2005;58(6):533-6.

\bibitem{eilam2014abn}
Eilam-Stock T, Xu P, Cao M, Gu X, Van~Dam NT, Anagnostou E, et~al.
\newblock Abnormal autonomic and associated brain activities during rest in autism spectrum disorder.
\newblock Brain. 2014;137(1):153-71.

\bibitem{neuhaus2016chi}
Neuhaus E, Bernier RA, Beauchaine TP.
\newblock Children with autism show altered autonomic adaptation to novel and familiar social partners.
\newblock Autism Research. 2016;9(5):579-91.

\bibitem{ellenbroek2017aut}
Ellenbroek BA, Sengul HK.
\newblock Autism spectrum disorders: Autonomic alterations with a special focus on the heart.
\newblock Heart and Mind. 2017;1(2):78-83.

\bibitem{condy2017res}
Condy EE, Scarpa A, Friedman BH.
\newblock Respiratory sinus arrhythmia predicts restricted repetitive behavior severity.
\newblock Journal of Autism and Developmental Disorders. 2017;47:2795-804.

\bibitem{theoharides2008nov}
Theoharides TC, Doyle R, Francis K, Conti P, Kalogeromitros D.
\newblock Novel therapeutic targets for autism.
\newblock Trends in Pharmacological Sciences. 2008;29(8):375-82.

\bibitem{gesundheit2013imm}
Gesundheit B, Rosenzweig JP, Naor D, Lerer B, Zachor DA, Proch{\'a}zka V, et~al.
\newblock Immunological and autoimmune considerations of Autism Spectrum Disorders.
\newblock Journal of autoimmunity. 2013;44:1-7.

\bibitem{dai2019inc}
Dai YX, Tai YH, Chang YT, Chen TJ, Chen MH.
\newblock Increased risk of atopic diseases in the siblings of patients with autism spectrum disorder: a nationwide population-based cohort study.
\newblock Journal of Autism and Developmental Disorders. 2019;49:4626-33.

\bibitem{wang2011pre}
Wang LW, Tancredi DJ, Thomas DW.
\newblock The prevalence of gastrointestinal problems in children across the United States with autism spectrum disorders from families with multiple affected members.
\newblock Journal of Developmental \& Behavioral Pediatrics. 2011;32(5):351-60.

\bibitem{sadik2022par}
Sadik A, Dardani C, Pagoni P, Havdahl A, Stergiakouli E, iPSYCH Autism Spectrum Disorder Working Group Grove Jakob 14 15 16~17, et~al.
\newblock Parental inflammatory bowel disease and autism in children.
\newblock Nature Medicine. 2022;28(7):1406-11.

\bibitem{wallon2008cor}
Keita {\AA}V.
\newblock Corticotropin-releasing hormone ({CRH}) regulates macromolecular permeability via mast cells in normal human colonic biopsies in vitro.
\newblock Gut. 2008;57(1):50-8.

\bibitem{kozicz2004uro}
Kozicz T, Korosi A, Korsman C, Groenink L, Veening J, van Der~Gugten J, et~al.
\newblock Urocortin expression in the {E}dinger-{W}estphal nucleus is down-regulated in transgenic mice over-expressing neuronal corticotropin-releasing factor.
\newblock Neuroscience. 2004;123(3):589-94.

\bibitem{reaven2011tre}
Reaven J.
\newblock The treatment of anxiety symptoms in youth with high-functioning autism spectrum disorders: Developmental considerations for parents.
\newblock Brain research. 2011;1380:255-63.

\bibitem{roberts2014wom}
Roberts AL, Koenen KC, Lyall K, Ascherio A, Weisskopf MG.
\newblock Women's posttraumatic stress symptoms and autism spectrum disorder in their children.
\newblock Research in autism spectrum disorders. 2014;8(6):608-16.

\bibitem{marinovic2008slo}
Marinovi{\'c}-{\'C}urin J, Marinovi{\'c}-Terzi{\'c} I, Bujas-Petkovi{\'c} Z, Zekan L, {\v{S}}krabi{\'c} V, {\DJ}oga{\v{s}} Z, et~al.
\newblock Slower cortisol response during {ACTH} stimulation test in autistic children.
\newblock European child \& adolescent psychiatry. 2008;17:39-43.

\bibitem{yang2015cor}
Yang CJ, Tan HP, Yang FY, Wang HP, Liu CL, He HZ, et~al.
\newblock The cortisol, serotonin and oxytocin are associated with repetitive behavior in autism spectrum disorder.
\newblock Research in Autism Spectrum Disorders. 2015;18:12-20.

\bibitem{varala2021dia}
Varala S, George R, Mathew L, Russell P, Koshy B, Oommen SP, et~al.
\newblock The diagnostic value of congenital and nevoid cutaneous lesions associated with autism Spectrum disorders in Indian children-a case-control study.
\newblock Indian Dermatology Online Journal. 2021;12(1):84-9.

\bibitem{sandru2019hyp}
Sandru F, Dumitrascu MC, Albu SE, Carsote M, Valea A.
\newblock Hyperpigmentation and {ACTH}--an overview of literature.
\newblock Ro Med J. 2019;66(4):309-12.

\bibitem{hirvikoski2016pre}
Hirvikoski T, Mittendorfer-Rutz E, Boman M, Larsson H, Lichtenstein P, B{\"o}lte S.
\newblock Premature mortality in autism spectrum disorder.
\newblock The British Journal of Psychiatry. 2016;208(3):232-8.

\bibitem{raghavan2018fet}
Raghavan R, Zuckerman B, Hong X, Wang G, Ji Y, Paige D, et~al.
\newblock Fetal and infancy growth pattern, cord and early childhood plasma leptin, and development of autism spectrum disorder in the {B}oston birth cohort.
\newblock Autism Research. 2018;11(10):1416-31.

\bibitem{yalnizoglu2007neu}
Yalnizoglu D, Haliloglu G, Turanli G, Cila A, Topcu M.
\newblock Neurologic outcome in patients with {MRI} pattern of damage typical for neonatal hypoglycemia.
\newblock Brain and Development. 2007;29(5):285-92.

\bibitem{wadhwa2004pla}
Wadhwa PD, Garite TJ, Porto M, Glynn L, Chicz-DeMet A, Dunkel-Schetter C, et~al.
\newblock Placental corticotropin-releasing hormone ({CRH}), spontaneous preterm birth, and fetal growth restriction: a prospective investigation.
\newblock American journal of obstetrics and gynecology. 2004;191(4):1063-9.

\bibitem{imperatore2006uro}
Imperatore A, Florio P, Torres PB, Torricelli M, Galleri L, Toti P, et~al.
\newblock Urocortin 2 and urocortin 3 are expressed by the human placenta, deciduas, and fetal membranes.
\newblock American journal of obstetrics and gynecology. 2006;195(1):288-95.

\bibitem{rubenstein2003mod}
Rubenstein J, Merzenich MM.
\newblock Model of autism: increased ratio of excitation/inhibition in key neural systems.
\newblock Genes, Brain and Behavior. 2003;2(5):255-67.

\bibitem{casanova2015min}
Casanova MF.
\newblock The minicolumnopathy of autism.
\newblock In: Recent Advances on the Modular Organization of the Cortex. Springer; 2015. p. 225-37.

\bibitem{courchesne2001unu}
Karns C, Davis H, Ziccardi R, Carper R, Tigue Z, Chisum H, et~al.
\newblock Unusual brain growth patterns in early life in patients with autistic disorder: an {MRI} study.
\newblock Neurology. 2001;57(2):245-54.

\bibitem{courchesne2003evi}
Courchesne E, Carper R, Akshoomoff N.
\newblock Evidence of brain overgrowth in the first year of life in autism.
\newblock Jama. 2003;290(3):337-44.

\bibitem{mills2007ele}
Mills JL, Hediger ML, Molloy CA, Chrousos GP, Manning-Courtney P, Yu KF, et~al.
\newblock Elevated levels of growth-related hormones in autism and autism spectrum disorder.
\newblock Clinical endocrinology. 2007;67(2):230-7.

\bibitem{chawarska2011ear}
Chawarska K, Campbell D, Chen L, Shic F, Klin A, Chang J.
\newblock Early generalized overgrowth in boys with autism.
\newblock Archives of general psychiatry. 2011;68(10):1021-31.

\bibitem{green2018ske}
Green CC, Dissanayake C, Loesch DZ, Bui M, Barbaro J.
\newblock Skeletal growth dysregulation in australian male infants and toddlers with autism spectrum disorder.
\newblock Autism Research. 2018;11(6):846-56.

\bibitem{dinstein2017no}
Dinstein I, Haar S, Atsmon S, Schtaerman H.
\newblock No evidence of early head circumference enlargements in children later diagnosed with autism in Israel.
\newblock Molecular autism. 2017;8(1):15.

\bibitem{zhong2005mul}
Zhong Q, Sridhar S, Ruan L, Ding KH, Xie D, Insogna K, et~al.
\newblock Multiple melanocortin receptors are expressed in bone cells.
\newblock Bone. 2005;36(5):820-31.

\bibitem{belmonte2004aut}
Belmonte MK, Allen G, Beckel-Mitchener A, Boulanger LM, Carper RA, Webb SJ.
\newblock Autism and abnormal development of brain connectivity.
\newblock J Neurosci. 2004;24(42):9228-31.

\bibitem{conti2015fir}
Conti E, Calderoni S, Marchi V, Muratori F, Cioni G, Guzzetta A.
\newblock The first 1000 days of the autistic brain: a systematic review of diffusion imaging studies.
\newblock Frontiers in human neuroscience. 2015;9:159.

\bibitem{pillion2018aud}
Pillion JP, Boatman-Reich D, Gordon B.
\newblock Auditory brainstem pathology in autism spectrum disorder: a review.
\newblock Cognitive and Behavioral Neurology. 2018;31(2):53-78.

\bibitem{kaplan2024mat}
Kaplan ZB, Pearce EN, Lee SY, Shin HM, Schmidt RJ.
\newblock Maternal Thyroid Dysfunction During Pregnancy as an Etiologic Factor in Autism Spectrum Disorder: Challenges and Opportunities for Research.
\newblock Thyroid. 2024;34(2):144-57.

\bibitem{rowland2021ass}
Rowland J, Wilson CA.
\newblock The association between gestational diabetes and {ASD} and {ADHD}: a systematic review and meta-analysis.
\newblock Scientific reports. 2021;11(1):5136.

\bibitem{chen2022type}
Chen MH, Tsai SJ, Bai YM, Huang KL, Su TP, Chen TJ, et~al.
\newblock Type 1 diabetes mellitus and risks of major psychiatric disorders: A nationwide population-based cohort study.
\newblock Diabetes \& Metabolism. 2022;48(1):101319.

\bibitem{liu2021poor}
Liu S, Kuja-Halkola R, Larsson H, Lichtenstein P, Ludvigsson JF, Svensson AM, et~al.
\newblock Poor glycaemic control is associated with increased risk of neurodevelopmental disorders in childhood-onset type 1 diabetes: a population-based cohort study.
\newblock Diabetologia. 2021;64(4):767-77.

\bibitem{manco2021cro}
Manco M, Guerrera S, Rav{\`a} L, Ciofi~degli Atti M, Di~Vara S, Valeri G, et~al.
\newblock Cross-sectional investigation of insulin resistance in youths with autism spectrum disorder. Any role for reduced brain glucose metabolism?
\newblock Translational Psychiatry. 2021;11(1):229.

\bibitem{sheikh2010bdn}
Sheikh AM, Malik M, Wen G, Chauhan A, Chauhan V, Gong CX, et~al.
\newblock {BDNF-A}kt-Bcl2 antiapoptotic signaling pathway is compromised in the brain of autistic subjects.
\newblock Journal of neuroscience research. 2010;88(12):2641-7.

\bibitem{tang2014los}
Tang G, Gudsnuk K, Kuo SH, Cotrina ML, Rosoklija G, Sosunov A, et~al.
\newblock Loss of m{TOR}-dependent macroautophagy causes autistic-like synaptic pruning deficits.
\newblock Neuron. 2014;83(5):1131-43.

\bibitem{nicolini2015dec}
Nicolini C, Ahn Y, Michalski B, Rho JM, Fahnestock M.
\newblock Decreased m{TOR} signaling pathway in human idiopathic autism and in rats exposed to valproic acid.
\newblock Acta neuropathologica communications. 2015;3:1-13.

\bibitem{russo2015dec}
Russo AJ.
\newblock Decreased phosphorylated protein kinase {B} ({A}kt) in individuals with autism associated with high epidermal growth factor receptor ({EGFR}) and low gamma-aminobutyric acid ({GABA}).
\newblock Biomarker insights. 2015;10:BMI-S21946.

\bibitem{gazestani2019per}
Gazestani VH, Pramparo T, Nalabolu S, Kellman BP, Murray S, Lopez L, et~al.
\newblock A perturbed gene network containing {PI3K--AKT, RAS--ERK} and {WNT}--$\beta$-catenin pathways in leukocytes is linked to {ASD} genetics and symptom severity.
\newblock Nature neuroscience. 2019;22(10):1624-34.

\bibitem{bjorklund2020oxi}
Bj{\o}rklund G, Meguid NA, El-Bana MA, Tinkov AA, Saad K, Dadar M, et~al.
\newblock Oxidative stress in autism spectrum disorder.
\newblock Molecular neurobiology. 2020;57:2314-32.

\bibitem{maimburg2008neo}
Maimburg RD, Vaeth M, Schendel DE, Bech BH, Olsen J, Thorsen P.
\newblock Neonatal jaundice: a risk factor for infantile autism?
\newblock Paediatric and perinatal epidemiology. 2008;22(6):562-8.

\bibitem{kujabi2021neo}
Kujabi ML, Petersen JP, Pedersen MV, Parner ET, Henriksen TB.
\newblock Neonatal jaundice and autism spectrum disorder: a systematic review and meta-analysis.
\newblock Pediatric Research. 2021;90(5):934-49.

\bibitem{lee2022glu}
Lee HY, Ithnin A, Azma RZ, Othman A, Salvador A, Cheah FC.
\newblock Glucose-6-Phosphate dehydrogenase deficiency and neonatal hyperbilirubinemia: Insights on pathophysiology, diagnosis, and gene variants in disease heterogeneity.
\newblock Frontiers in Pediatrics. 2022;10:875877.

\bibitem{majewska2014mar}
Majewska MD, Hill M, Urbanowicz E, Rok-Bujko P, Bie{\'n}kowski P, Namys{\l}owska I, et~al.
\newblock Marked elevation of adrenal steroids, especially androgens, in saliva of prepubertal autistic children.
\newblock European child \& adolescent psychiatry. 2014;23:485-98.

\bibitem{elBaz2014hyp}
El-Baz F, Hamza RT, Ayad MS, Mahmoud NH.
\newblock Hyperandrogenemia in male autistic children and adolescents: relation to disease severity.
\newblock International journal of adolescent medicine and health. 2014;26(1):79-84.

\bibitem{hassan2019pos}
Hassan MH, Desoky T, Sakhr HM, Gabra RH, Bakri AH.
\newblock Possible metabolic alterations among autistic male children: clinical and biochemical approaches.
\newblock Journal of Molecular Neuroscience. 2019;67:204-16.

\bibitem{janvsakova2020alt}
Jan{\v{s}}{\'a}kov{\'a} K, Hill M, {\v{C}}el{\'a}rov{\'a} D, Celu{\v{s}}{\'a}kov{\'a} H, Repisk{\'a} G, Bi{\v{c}}{\'\i}kov{\'a} M, et~al.
\newblock Alteration of the steroidogenesis in boys with autism spectrum disorders.
\newblock Translational Psychiatry. 2020;10(1):340.

\bibitem{alZaid2021pot}
Al-Zaid FS, Alhader AFA, Al-Ayadhi LY.
\newblock A potential role for the adrenal gland in autism.
\newblock Scientific Reports. 2021;11(1):17743.

\bibitem{geier2007pro}
Geier DA, Geier MR.
\newblock A prospective assessment of androgen levels in patients with autistic spectrum disorders: biochemical underpinnings and suggested therapies.
\newblock Neuroendocrinology Letters. 2007;28(5):565-74.

\bibitem{schmidtova2010pol}
Schmidtova E, Kelemenova S, Celec P, Ficek A, Ostatnikova D.
\newblock Polymorphisms in genes involved in testosterone metabolism in {S}lovak autistic boys.
\newblock The Endocrinologist. 2010;20(5):245-9.

\bibitem{ruta2011inc}
Ruta L, Ingudomnukul E, Taylor K, Chakrabarti B, Baron-Cohen S.
\newblock Increased serum androstenedione in adults with autism spectrum conditions.
\newblock Psychoneuroendocrinology. 2011;36(8):1154-63.

\bibitem{alZaid2014alt}
Al-Zaid FS, Alhader AA, Al-Ayadhi LY.
\newblock Altered ghrelin levels in boys with autism: a novel finding associated with hormonal dysregulation.
\newblock Scientific reports. 2014;4(1):6478.

\bibitem{ostatnikova2016neu}
Ostatnikova D, Kubranska A, Marcincakova V, Pivovarciova A, Babkova-Durdiakova J.
\newblock Neuroendocrine contribution to autism etiology.
\newblock Act Nerv Super Rediviva. 2016;58(3):65-8.

\bibitem{procyshyn2020eff}
Procyshyn TL, Lombardo MV, Lai MC, Auyeung B, Crockford SK, Baron-Cohen S, et~al.
\newblock Effects of oxytocin administration on salivary sex hormone levels in autistic and neurotypical women.
\newblock Molecular Autism. 2020;11:1-11.

\bibitem{muscatello2022sal}
Muscatello RA, Rafatjoo E, Mirpuri KK, Kim A, Vandekar S, Corbett BA.
\newblock Salivary testosterone in male and female youth with and without autism spectrum disorder: considerations of development, sex, and diagnosis.
\newblock Molecular Autism. 2022;13(1):37.

\bibitem{croonenberghs2010ser}
Croonenberghs J, Van~Grieken S, Wauters A, Van~West D, Brouw L, Maes M, et~al.
\newblock Serum testosterone concentration in male autistic youngsters.
\newblock Neuroendocrinology Letters. 2010;31(4):483.

\bibitem{krajmer2011rel}
Krajmer P, Spajdel M, Celec P, Ostatn{\'\i}kov{\'a} D.
\newblock Relationship between salivary testosterone levels and empathizing/systemizing in {S}lovak boys with {A}sperger syndrome.
\newblock Studia Psychologica. 2011;53(3):293.

\bibitem{sharpley2017age}
Sharpley CF, Bitsika V, Andronicos NM, Agnew LL.
\newblock Age-Related Variations in Comparative Testosterone Concentrations Between Boys with Autism Spectrum Disorder and their typically-Developing Peers: A Challenge to the 'Extreme Male Brain' Hypothesis of {ASD}.
\newblock Journal of Developmental and Physical Disabilities. 2017;29(2):353-67.

\bibitem{sriram2022eva}
Sriram N, Madaan P, Malhi P, Sachdeva N, Negi S, Das J, et~al.
\newblock Evaluation of Hyperandrogenism in Children with Autism Spectrum Disorder.
\newblock Indian Journal of Pediatrics. 2022;89(7):717-9.

\bibitem{baron2020foe}
Baron-Cohen S, Tsompanidis A, Auyeung B, N{\o}rgaard-Pedersen B, Hougaard DM, Abdallah M, et~al.
\newblock Foetal oestrogens and autism.
\newblock Molecular Psychiatry. 2020;25(11):2970-8.

\bibitem{whitehouse2012per}
Whitehouse AJ, Mattes E, Maybery MT, Dissanayake C, Sawyer M, Jones RM, et~al.
\newblock Perinatal testosterone exposure and autistic-like traits in the general population: a longitudinal pregnancy-cohort study.
\newblock Journal of neurodevelopmental disorders. 2012;4:1-12.

\bibitem{kung2016traits}
Kung KT, Spencer D, Pasterski V, Neufeld S, Glover V, O'Connor TG, et~al.
\newblock No relationship between prenatal androgen exposure and autistic traits: convergent evidence from studies of children with congenital adrenal hyperplasia and of amniotic testosterone concentrations in typically developing children.
\newblock Journal of Child Psychology and Psychiatry. 2016;57(12):1455-62.

\bibitem{auyeung2012pre}
Auyeung B, Ahluwalia J, Thomson L, Taylor K, Hackett G, O’Donnell KJ, et~al.
\newblock Prenatal versus postnatal sex steroid hormone effects on autistic traits in children at 18 to 24 months of age.
\newblock Molecular autism. 2012;3(1):17.

\bibitem{saenz2013pos}
Saenz J, Alexander GM.
\newblock Postnatal testosterone levels and disorder relevant behavior in the second year of life.
\newblock Biological psychology. 2013;94(1):152-9.

\bibitem{kung2016no}
Kung KT, Constantinescu M, Browne WV, Noorderhaven RM, Hines M.
\newblock No relationship between early postnatal testosterone concentrations and autistic traits in 18 to 30-month-old children.
\newblock Molecular autism. 2016;7(1):15.

\bibitem{metwally2020imp}
Metwally FM, Rashad H, Zeidan HM, Hashish AF.
\newblock Impact of bisphenol {A} on gonadotropic hormone levels in children with autism spectrum disorders.
\newblock Indian Journal of Clinical Biochemistry. 2020;35:205-10.

\bibitem{kosidou2016mat}
Kosidou K, Dalman C, Widman L, Arver S, Lee B, Magnusson C, et~al.
\newblock Maternal polycystic ovary syndrome and the risk of autism spectrum disorders in the offspring: a population-based nationwide study in {S}weden.
\newblock Molecular psychiatry. 2016;21(10):1441.

\bibitem{rotem2021ass}
Rotem RS, Nguyen VT, Chodick G, Davidovitch M, Shalev V, Hauser R, et~al.
\newblock Associations of maternal androgen-related conditions with risk of autism spectrum disorder in progeny and mediation by cardiovascular, metabolic, and fertility factors.
\newblock American journal of epidemiology. 2021;190(4):600-10.

\bibitem{xu2013mot}
Xu XJ, Shou XJ, Li J, Jia MX, Zhang JS, Guo Y, et~al.
\newblock Mothers of autistic children: lower plasma levels of oxytocin and Arg-vasopressin and a higher level of testosterone.
\newblock PLoS One. 2013;8(9):e74849.

\bibitem{varcin2017pre}
Varcin KJ, Alvares GA, Uljarevi{\'c} M, Whitehouse AJ.
\newblock Prenatal maternal stress events and phenotypic outcomes in Autism Spectrum Disorder.
\newblock Autism Research. 2017;10(11):1866-77.

\bibitem{seebeck2023ass}
Seebeck J, Sznajder KK, Kjerulff KH.
\newblock The association between prenatal psychosocial factors and autism spectrum disorder in offspring at 3 years: a prospective cohort study.
\newblock Social psychiatry and psychiatric epidemiology. 2023:1-11.

\bibitem{atladottir2010mat}
Atlad{\'o}ttir H{\'O}, Thorsen P, {\O}stergaard L, Schendel DE, Lemcke S, Abdallah M, et~al.
\newblock Maternal infection requiring hospitalization during pregnancy and autism spectrum disorders.
\newblock Journal of autism and developmental disorders. 2010;40:1423-30.

\bibitem{walder2014pre}
Walder DJ, Laplante DP, Sousa-Pires A, Veru F, Brunet A, King S.
\newblock Prenatal maternal stress predicts autism traits in 6$1/2$ year-old children: Project Ice Storm.
\newblock Psychiatry research. 2014;219(2):353-60.

\bibitem{barrett2013pre}
Barrett ES, Parlett LE, Sathyanarayana S, Liu F, Redmon JB, Wang C, et~al.
\newblock Prenatal exposure to stressful life events is associated with masculinized anogenital distance ({AGD}) in female infants.
\newblock Physiology \& behavior. 2013;114:14-20.

\bibitem{glidden2016gen}
Glidden D, Bouman WP, Jones BA, Arcelus J.
\newblock Gender dysphoria and autism spectrum disorder: A systematic review of the literature.
\newblock Sexual Medicine Reviews. 2016;4(1):3-14.

\bibitem{weir2021sex}
Weir E, Allison C, Baron-Cohen S.
\newblock The sexual health, orientation, and activity of autistic adolescents and adults.
\newblock Autism Research. 2021;14(11):2342-54.

\bibitem{rudolph2018bri}
Rudolph CE, Lundin A, {\AA}hs JW, Dalman C, Kosidou K.
\newblock Brief report: Sexual orientation in individuals with autistic traits: Population based study of 47,000 adults in {S}tockholm County.
\newblock Journal of Autism and Developmental Disorders. 2018;48:619-24.

\bibitem{bejerot2012ext}
Bejerot S, Eriksson JM, Bonde S, Carlstr{\"o}m K, Humble MB, Eriksson E.
\newblock The extreme male brain revisited: gender coherence in adults with autism spectrum disorder.
\newblock The British Journal of Psychiatry. 2012;201(2):116-23.

\bibitem{gilani2015sex}
Gilani SZ, Tan DW, Russell-Smith SN, Maybery MT, Mian A, Eastwood PR, et~al.
\newblock Sexually dimorphic facial features vary according to level of autistic-like traits in the general population.
\newblock Journal of neurodevelopmental disorders. 2015;7(1):14.

\bibitem{tan2017hyp}
Tan DW, Gilani SZ, Maybery MT, Mian A, Hunt A, Walters M, et~al.
\newblock Hypermasculinised facial morphology in boys and girls with Autism Spectrum Disorder and its association with symptomatology.
\newblock Scientific Reports. 2017;7(1):9348.

\bibitem{calogero1996eff}
Calogero AE, Burrello N, Negri-Cesi P, Papale L, Palumbo MA, Cianci A, et~al.
\newblock Effects of corticotropin-releasing hormone on ovarian estrogen production in vitro.
\newblock Endocrinology. 1996;137(10):4161-6.

\bibitem{deCock2012doe}
De~Cock M, Maas YG, Van De~Bor M.
\newblock Does perinatal exposure to endocrine disruptors induce autism spectrum and attention deficit hyperactivity disorders? Review.
\newblock Acta paediatrica. 2012;101(8):811-8.

\bibitem{hansen2021pre}
Hansen JB, Bilenberg N, Timmermann CAG, Jensen RC, Frederiksen H, Andersson AM, et~al.
\newblock Prenatal exposure to bisphenol {A} and autistic-and {ADHD}-related symptoms in children aged 2 and5 years from the {O}dense Child Cohort.
\newblock Environmental Health. 2021;20:1-12.

\bibitem{phumsatitpong2018est}
Phumsatitpong C, Moenter SM.
\newblock Estradiol-dependent stimulation and suppression of gonadotropin-releasing hormone neuron firing activity by corticotropin-releasing hormone in female mice.
\newblock Endocrinology. 2018;159(1):414-25.

\bibitem{schottle2017sex}
Sch{\"o}ttle D, Briken P, T{\"u}scher O, Turner D.
\newblock Sexuality in autism: hypersexual and paraphilic behavior in women and men with high-functioning autism spectrum disorder.
\newblock Dialogues in clinical neuroscience. 2017;19(4):381-93.

\bibitem{feldman2014par}
Feldman R, Golan O, Hirschler-Guttenberg Y, Ostfeld-Etzion S, Zagoory-Sharon O.
\newblock Parent--child interaction and oxytocin production in pre-schoolers with autism spectrum disorder.
\newblock The British Journal of Psychiatry. 2014;205(2):107-12.

\bibitem{john2021oxy}
John S, Jaeggi AV.
\newblock Oxytocin levels tend to be lower in autistic children: A meta-analysis of 31 studies.
\newblock Autism. 2021;25(8):2152-61.

\bibitem{yang2017ser}
Yang S, Dong X, Guo X, Han Y, Song H, Gao L, et~al.
\newblock Serum oxytocin levels and an oxytocin receptor gene polymorphism (rs2254298) indicate social deficits in children and adolescents with autism spectrum disorders.
\newblock Frontiers in neuroscience. 2017;11:230653.

\bibitem{phaneuf1997des}
Phaneuf S, Asboth G, Carrasco M, Europe-Finner G, Saji F, Kimura T, et~al.
\newblock The desensitization of oxytocin receptors in human myometrial cells is accompanied by down-regulation of oxytocin receptor messenger {RNA}.
\newblock Journal of Endocrinology. 1997;154(1):7-18.

\bibitem{friedlander2017soc}
Friedlander E, Feldstein O, Mankuta D, Yaari M, Harel-Gadassi A, Ebstein RP, et~al.
\newblock Social impairments among children perinatally exposed to oxytocin or oxytocin receptor antagonist.
\newblock Early Human Development. 2017;106:13-8.

\bibitem{boucher2017ass}
Boucher O, Julvez J, Guxens M, Arranz E, Ibarluzea J, Sanchez~de Miguel M, et~al.
\newblock Association between breastfeeding duration and cognitive development, autistic traits and {ADHD} symptoms: a multicenter study in {S}pain.
\newblock Pediatric Research. 2017;81(3):434-42.

\bibitem{tseng2019mat}
Tseng PT, Chen YW, Stubbs B, Carvalho AF, Whiteley P, Tang CH, et~al.
\newblock Maternal breastfeeding and autism spectrum disorder in children: A systematic review and meta-analysis.
\newblock Nutritional neuroscience. 2019;22(5):354-62.

\bibitem{soke2019ass}
Soke GN, Maenner M, Windham G, Moody E, Kaczaniuk J, DiGuiseppi C, et~al.
\newblock Association between breastfeeding initiation and duration and autism spectrum disorder in preschool children enrolled in the study to explore early development.
\newblock Autism Research. 2019;12(5):816-29.

\bibitem{chen2021ass}
Chen J, Strodl E, Huang LH, Chen JY, Liu XC, Yang JH, et~al.
\newblock Associations between prenatal education, breastfeeding and autistic-like behaviors in pre-schoolers.
\newblock Children. 2021;8(2):124.

\bibitem{abd2022bre}
Abd~Elmaksoud MS, Aly O, Abd~Elfatah M, Mahfouz A.
\newblock Breastfeeding and autism spectrum disorder: a cross-sectional study from {E}gypt.
\newblock Alexandria Journal of Pediatrics. 2022;35(1):59-66.

\bibitem{xiang2023ass}
Xiang X, Yang T, Chen J, Chen L, Dai Y, Zhang J, et~al.
\newblock Association of feeding patterns in infancy with later autism symptoms and neurodevelopment: a national multicentre survey.
\newblock BMC psychiatry. 2023;23(1):174.

\bibitem{zhan2023inf}
Zhan XL, Pan N, Karatela S, Shi L, Wang X, Liu ZY, et~al.
\newblock Infant feeding practices and autism spectrum disorder in {US} children aged 2--5 years: the national survey of children’s health ({NSCH}) 2016--2020.
\newblock International breastfeeding journal. 2023;18(1):41.

\bibitem{plotsky1993ear}
Plotsky PM, Meaney MJ.
\newblock Early, postnatal experience alters hypothalamic corticotropin-releasing factor ({CRF}) m{RNA}, median eminence {CRF} content and stress-induced release in adult rats.
\newblock Molecular brain research. 1993;18(3):195-200.

\bibitem{avishai2001down}
Avishai-Eliner S, Eghbal-Ahmadi M, Tabachnik E, Brunson KL, Baram TZ.
\newblock Down-regulation of hypothalamic corticotropin-releasing hormone messenger ribonucleic acid (m{RNA}) precedes early-life experience-induced changes in hippocampal glucocorticoid receptor m{RNA}.
\newblock Endocrinology. 2001;142(1):89-97.

\bibitem{kaiser2015bra}
Kaiser MD, Yang DYJ, Voos AC, Bennett RH, Gordon I, Pretzsch C, et~al.
\newblock Brain mechanisms for processing affective (and nonaffective) touch are atypical in autism.
\newblock Cerebral Cortex. 2016;26(6):2705-14.

\bibitem{lucas2015dys}
Lucas RF, Cutler A.
\newblock Dysregulated breastfeeding behaviors in children later diagnosed with autism.
\newblock The Journal of Perinatal Education. 2015;24(3):171.

\bibitem{kara2022nur}
Kara T, Alpgan {\"O}.
\newblock Nursing personality and features in children with autism spectrum disorder aged 0--2: an exploratory case-control study.
\newblock Nutritional neuroscience. 2022;25(6):1200-8.

\bibitem{peries2023bre}
Peries M, Duhr F, Picot MC, Heude B, Bernard JY, Baghdadli A.
\newblock Breastfeeding is not a risk factor for clinical severity in Autism spectrum disorder in children from the {ELENA} cohort.
\newblock Scientific Reports. 2023;13(1):816.

\bibitem{varma2023fee}
Varma C, de~Souza N.
\newblock Feeding behaviours in infancy of children later diagnosed with autism spectrum disorder.
\newblock International Journal of Contemporary Pediatrics. 2023;10(8):1280.

\bibitem{vantHof2021ear}
van't Hof M, Ester WA, van Berckelaer-Onnes I, Hillegers MH, Hoek HW, Jansen PW.
\newblock Do early-life eating habits predict later autistic traits? Results from a population-based study.
\newblock Appetite. 2021;156:104976.

\bibitem{klockars2015cen}
Klockars A, Levine AS, Olszewski PK.
\newblock Central oxytocin and food intake: focus on macronutrient-driven reward.
\newblock Frontiers in endocrinology. 2015;6:136509.

\bibitem{wallace2018aut}
Wallace GL, Llewellyn C, Fildes A, Ronald A.
\newblock Autism spectrum disorder and food neophobia: clinical and subclinical links.
\newblock The American journal of clinical nutrition. 2018;108(4):701-7.

\bibitem{yabut2019eme}
Yabut JM, Crane JD, Green AE, Keating DJ, Khan WI, Steinberg GR.
\newblock Emerging roles for serotonin in regulating metabolism: new implications for an ancient molecule.
\newblock Endocrine reviews. 2019;40(4):1092-107.

\bibitem{muller2016ser}
Muller CL, Anacker AM, Veenstra-VanderWeele J.
\newblock The serotonin system in autism spectrum disorder: From biomarker to animal models.
\newblock Neuroscience. 2016;321:24-41.

\bibitem{lin2024ris}
Lin J, Costa MA, Rezende VL, Nascimento RR, Ambr{\'o}sio PG, Madeira K, et~al.
\newblock Risk factors and clinical profile of autism spectrum disorder in southern {B}razil.
\newblock Journal of Psychiatric Research. 2024;169:105-12.

\bibitem{lazo1960pit}
Lazo-Wasem E.
\newblock Pituitary {ACTH} levels during adrenal involution following thiouracil.
\newblock Proceedings of the Society for Experimental Biology and Medicine. 1960;103(2):300-2.

\bibitem{miskowiak1986com}
Mi{\'s}kowiak B, Kasprzak A, Malendowicz L.
\newblock Comparative stereological studies on the effects of long term {CRF} and {ACTH} treatment on the cortex of the suprarenal gland.
\newblock Journal of anatomy. 1986;146:167.

\bibitem{hannerfors2015tre}
Hannerfors AK, Hellgren C, Schijven D, Iliadis SI, Comasco E, Skalkidou A, et~al.
\newblock Treatment with serotonin reuptake inhibitors during pregnancy is associated with elevated corticotropin-releasing hormone levels.
\newblock Psychoneuroendocrinology. 2015;58:104-13.

\bibitem{sandbank2020pro}
Sandbank M, Bottema-Beutel K, Crowley S, Cassidy M, Dunham K, Feldman JI, et~al.
\newblock Project {AIM}: Autism intervention meta-analysis for studies of young children.
\newblock Psychological bulletin. 2020;146(1):1.

\bibitem{futch2019ant}
Futch HS, McFarland KN, Moore BD, Kuhn MZ, Giasson BI, Ladd TB, et~al.
\newblock An anti-{CRF} antibody suppresses the {HPA} axis and reverses stress-induced phenotypes.
\newblock Journal of Experimental Medicine. 2019;216(11):2479-91.

\bibitem{biason2013mar}
Biason-Lauber A, Miller WL, Pandey AV, Fl{\"u}ck CE.
\newblock Of marsupials and men: `Backdoor' dihydrotestosterone synthesis in male sexual differentiation.
\newblock Molecular and cellular endocrinology. 2013;371(1-2):124-32.

\bibitem{lund2006and}
Lund TD, Hinds LR, Handa RJ.
\newblock The androgen 5$\alpha$-dihydrotestosterone and its metabolite 5$\alpha$-androstan-3$\beta$, 17$\beta$-diol inhibit the hypothalamo--pituitary--adrenal response to stress by acting through estrogen receptor $\beta$-expressing neurons in the hypothalamus.
\newblock Journal of Neuroscience. 2006;26(5):1448-56.

\bibitem{handa2013cen}
Handa RJ, Kudwa AE, Donner NC, McGivern RF, Brown R.
\newblock Central 5-alpha reduction of testosterone is required for testosterone's inhibition of the hypothalamo-pituitary--adrenal axis response to restraint stress in adult male rats.
\newblock Brain research. 2013;1529:74-82.

\bibitem{kitay1971eff}
KITAY JI, COYNE MD, SWYGERT NH.
\newblock Effects of hypophysectomy and administration of cortisone or {ACTH} on adrenal 5$\alpha$-reductase activity and steroid production.
\newblock Endocrinology. 1971;89(2):432-8.

\bibitem{lee2020lab}
Lee M, Shnitko T, Blue S, Kaucher A, Winchell A, Erikson D, et~al.
\newblock Labeled oxytocin administered via the intranasal route reaches the brain in rhesus macaques.
\newblock Nature communications. 2020;11(1):2783.

\bibitem{pan2008uro}
Pan W, Kastin AJ.
\newblock Urocortin and the brain.
\newblock Progress in neurobiology. 2008;84(2):148-56.

\bibitem{yenkoyan2017adv}
Yenkoyan K, Grigoryan A, Fereshetyan K, Yepremyan D.
\newblock Advances in understanding the pathophysiology of autism spectrum disorders.
\newblock Behavioural brain research. 2017;331:92-101.

\bibitem{baron2010emp}
Baron-Cohen S.
\newblock Empathizing, systemizing, and the extreme male brain theory of autism.
\newblock Progress in brain research. 2010;186:167-75.

\bibitem{carter2007sex}
Carter CS.
\newblock Sex differences in oxytocin and vasopressin: implications for autism spectrum disorders?
\newblock Behavioural brain research. 2007;176(1):170-86.

\bibitem{dickinson2016mea}
Dickinson A, Jones M, Milne E.
\newblock Measuring neural excitation and inhibition in autism: Different approaches, different findings and different interpretations.
\newblock Brain research. 2016;1648:277-89.

\bibitem{courchesne2020pre}
Courchesne E, Gazestani VH, Lewis NE.
\newblock Prenatal origins of {ASD}: the when, what, and how of {ASD} development.
\newblock Trends in neurosciences. 2020;43(5):326-42.

\bibitem{wong2014met}
Wong C, Price T, Schalkwyk L, Plomin R, Mill J.
\newblock Methylomic analysis of monozygotic twins discordant for autism spectrum disorder and related behavioural traits.
\newblock Molecular psychiatry. 2014;19(4):495-503.

\bibitem{frasch2023aut}
Frasch MG, Yoon BJ, Helbing DL, Snir G, Antonelli MC, Bauer R.
\newblock Autism spectrum disorder: a neuro-immunometabolic hypothesis of the developmental origins.
\newblock Biology. 2023;12(7):914.

\end{thebibliography}

\end{document}